Review Article
# Integrating MEMS and ICs

Andreas C. Fischer, Fredrik Forsberg, Martin Lapisa, Simon J. Bleiker, Göran Stemme, Niclas Roxhed and Frank Niklaus

The majority of microelectromechanical system (MEMS) devices must be combined with integrated circuits (ICs) for operation in larger electronic systems. While MEMS transducers sense or control physical, optical or chemical quantities, ICs typically provide functionalities related to the signals of these transducers, such as analog-to-digital conversion, amplification, filtering and information processing as well as communication between the MEMS transducer and the outside world. Thus, the vast majority of commercial MEMS products, such as accelerometers, gyroscopes and micro-mirror arrays, are integrated and packaged together with ICs. There are a variety of possible methods of integrating and packaging MEMS and IC components, and the technology of choice strongly depends on the device, the field of application and the commercial requirements. In this review paper, traditional as well as innovative and emerging approaches to MEMS and IC integration are reviewed. These include approaches based on the hybrid integration of multiple chips (multi-chip solutions) as well as system-on-chip solutions based on wafer-level monolithic integration and heterogeneous integration techniques. These are important technological building blocks for the 'More-Than-Moore' paradigm described in the International Technology Roadmap for Semiconductors. In this paper, the various approaches are categorized in a coherent manner, their merits are discussed, and suitable application areas and implementations are critically investigated. The implications of the different MEMS and IC integration approaches for packaging, testing and final system costs are reviewed.

## 1 INTRODUCTION

Microelectromechanical system (MEMS) and emerging nanoelectromechanical system (NEMS), henceforth both referred to as 'MEMS', are typically transducer systems that sense or control physical, optical or chemical quantities, such as acceleration, radiation or fluids. A MEMS device typically interacts with a physical, chemical or optical quantity and has an electrical interface to the outside world. For MEMS sensors, the electrical output signal correlates with the physical, optical or chemical input quantity that is sensed. In the case of MEMS actuators, an electrical input signal is used to control one or more physical, optical or chemical quantities. To enable the MEMS transducer to perform useful functions, the electrical interface with the outside world is, in most cases, realized through integrated circuits (ICs) that provide the system with the necessary intelligence. ICs may provide signal conditioning functions such as analog-to-digital conversion, amplification, temperature compensation, storage or filtering as well as system testing and logic and communication functions. The International Technology Roadmap for Semiconductors[1] describes the addition of MEMS functionalities to ICs as an important 'More-Than-Moore' technology.

As illustrated in Figure 1, MEMS and ICs can be integrated using two basic methods: (1) In the general approach referred to here as a multi-chip solution, MEMS and IC components are manufactured on separate substrates using dedicated MEMS and IC processes and are subsequently hybridized in the final system. Two-dimensionally or side-by-side integrated systems are often referred to as multi-chip modules. When chips are vertically stacked in a package in this way, such a system is also referred to as a system-in-package or a vertical multi-chip module[2]. Devices created through vertical stacking of several IC chips are also referred to as three-dimensional integrated circuits (3D ICs)[3]. (2) In the general approach referred to here as a system-on-chip (SoC) solution, MEMS and IC components are manufactured on the same substrate, using consecutive or interlaced processing schemes.

For both multi-chip solutions and SoC solutions, numerous technological schemes have been proposed, and the research community and industry continue to develop new manufacturing and integration schemes at a rapid pace. Each of the two basic methods of combining MEMS and ICs offers distinct advantages and disadvantages, and the preferred solution depends strongly on the device, the field of application and the product requirements. Based on recent MEMS market studies[4,5], we estimate that approximately half of all existing MEMS products (in terms of market value) are currently implemented as multi-chip solutions (including many accelerometers, gyroscopes, microphones, pressure sensors, RF MEMS and microfluidic devices) and that the other half are implemented as SoC solutions (including digital mirror devices, infrared bolometer arrays, inkjet printheads, and certain gyroscopes, accelerometers and pressure sensors). Many of the MEMS products that are implemented as SoC solutions have the common feature that they consist of large transducer arrays in which each transducer is operated individually, and thus, the integration of each MEMS transducer and its associated IC on a single chip is the only practical way to implement these types of systems. However, there are also other products, including certain gyroscopes, accelerometers, microphones and pressure sensors, that are implemented as SoC solutions. The commonality among these products is that they are relatively mature products that are manufactured and sold in very high volumes.

[1]Department of Micro and Nanosystems, School of Electrical Engineering, KTH Royal Institute of Technology, SE-10044 Stockholm, Sweden
Correspondence: Andreas C. Fischer (andreas.fischer@ee.kth.se)

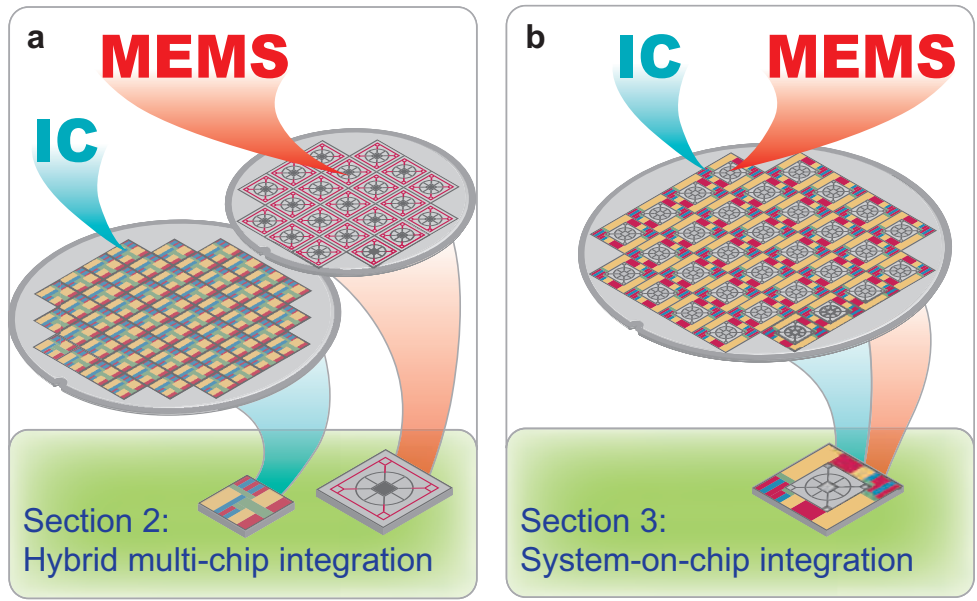

**Figure 1** MEMS and IC integration methods are based on either (**a**) hybrid multi-chip solutions (described in Section 2) or (**b**) system-on-chip solutions (described in Section 3).

The existing literature contains several reviews of important technologies for the cofabrication of MEMS and electronics on the same substrate (by Fedder et al[6], Brand[7] and French et al[8]). Several review articles on heterogeneous integration technologies for MEMS have also been published[9–12]. In contrast to previous work, the present paper provides a comprehensive, up-to-date overview and comparison of the available technologies for integrating MEMS and ICs, both for (1) hybrid multi-chip solutions and (2) SoC solutions. Established and emerging technologies for the integration of MEMS and ICs are reviewed, analyzed and categorized in a coherent manner, and their implementation in MEMS products is discussed. MEMS packages must provide either a hermetically sealed environment (e.g., for resonators, inertial sensors and IR imaging sensors) or a physical interface to the ambient environment (e.g., for microphones, pressure sensors and flow sensors). The packaging of MEMS is not the subject of the present review; however, this topic is discussed in a number of review articles[13–16] and in books by Hsu[17], Tummala[18], Lau[19] and Madou[20].

## 2 HYBRID INTEGRATION OF MEMS AND ICS: MULTI-CHIP SOLUTIONS

In recent decades, hybrid integration of MEMS and IC technology has been dominated by 2D integration approaches. In such an approach, the MEMS and IC wafers are designed, manufactured and tested independently. The wafers are then separated into discrete chips and eventually integrated into multi-chip systems at the board or package level. Historically, MEMS and IC chips have been packaged individually and then integrated as a system onto a printed circuit board (PCB). This led to the development of multi-chip modules in which, as shown in Figure 2, MEMS and IC chips are placed side-by-side in a common package and interconnected at the package level, typically via wire and/or flip-chip bonding. In wire bonding, a highly automated micro-welding process for metal wires is employed to create chip-to-chip and chip-to-package interconnects[21]. In flip-chip bonding, solder balls or stud bumps are placed on pads on the topside of a chip. The chip is then flipped upside down and aligned with and attached to the package substrate or another chip via pick-and-place soldering. Both the wire and flip-chip bonding processes employ temperature, force and/or ultrasonic energy in the joining process. While wire bonds serve merely as electrical interconnects, flip-chip bonds can double as physical chip attachments and electrical interconnects. Several concepts for MEMS and IC integration through flip-chip bonding are being used in applications such as RF-MEMS[22], micro-opto-electro-mechanical system[23–25] and MEMS sensors[26,27], and a general study investigating various MEMS test structures has been reported[28]. More recent concepts for chip-to-chip and chip-to-package interconnections have been developed based on thin-film interconnects between embedded chips[29–33]. This method has been commercially exploited in fan-out wafer-level packaging concepts, including the wafer-level ball grid arrays developed by Infineon and there distributed chip packages developed by Freescale Semiconductor Inc, Austin, Texas, USA[34]. Other unconventional chip-to-chip interconnection methods include mechanically flexible interconnects[35] and quilt packaging, in which chips with different functionalities are tiled in close proximity on the package substrate and are interconnected by vertical facets protruding from each chip[36].

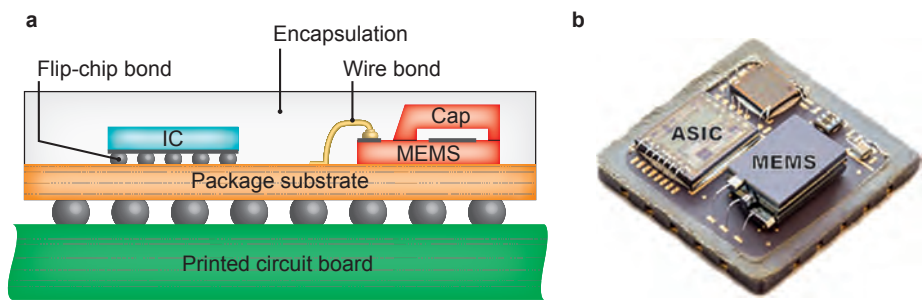

**Figure 2** (**a**) 2D side-by-side integration with flip-chip and wire bonded interconnections. (**b**) Photograph of a decapsulated Colyibrys MS9000-series accelerometer (Colibrys Ltd, Yverdon-les-Bains, Switzerland). ASIC and passive chips are placed side-by-side with an encapsulated MEMS chip. All chips have wire bonded interconnections. Package dimensions: $8.9 \times 8.9 \times 3.2$ mm$^3$.

Multi-chip modules have a significantly reduced signal path length between chips and occupy a lesser area of PCB real estate compared with system-on-board approaches, in which discrete components are integrated on a PCB. Thus, this concept is broadly applied both in research[37–42] and in commercial products[27,39]. In particular, the integration of MEMS chips with commercially available standard application-specific integrated circuits (ASICs) enables the extremely simple, rapid and cost-efficient implementation of hybrid systems[41].

As shown in Figure 3, system-in-packages, also referred to as vertical or stacked multi-chip modules, consist of chips that are attached on top of each other and interconnected via wire and/or flip-chip bonding, either directly[43,44] or through additional re-distribution layers[45]. The main benefits of these 3D-stacked approaches are their higher integration densities, shorter signal path lengths and smaller package footprints/volumes in comparison with multi-chip modules. This method enables chip-to-wafer stacking[43,44,46] and is employed both in research[47–50] and in a large number of commercial products, such as accelerometers and pressure sensors[51–53].

As shown in Figure 4, the chip-scale package and wafer-scale package concepts yield very compact packages with footprints similar to the size of the largest chip involved. An example of this approach is the chip-on-MEMS technology that has been commercialized by Murata Electronics Oy, Finland (formerly VTI Technology Oy, Vantaa, Finland). Here, IC chips are attached to larger, encapsulated MEMS devices at the wafer level via flip-chip bonding (chip-to-wafer bonding). After dicing, the chip-stack is again flip-chip bonded directly to the PCB, as depicted in Figure 4a. Similar approaches have been reported by Premachandran et al[55], Tian et al[56] and Sugizaki et al[57]. Such compact integration concepts require vertical through-substrate vias, known as through-silicon vias or through-glass vias, depending on the substrate used. Through-substrate vias enable even shorter signal path lengths with superior electrical characteristics in terms of lower capacitive, resistive and inductive parasitic effects. Through-substrate vias have already been commercially implemented in a number of products, such as MEMS-microphones[58] and RF-filter and RF-switch devices[59,60].

System-on-package approaches are another alternative that enables highly integrated and miniaturized system technology at the package level. Here, MEMS and IC devices are integrated with a broad spectrum of other basic technologies, ranging from optics and power electronics to wireless components, in a

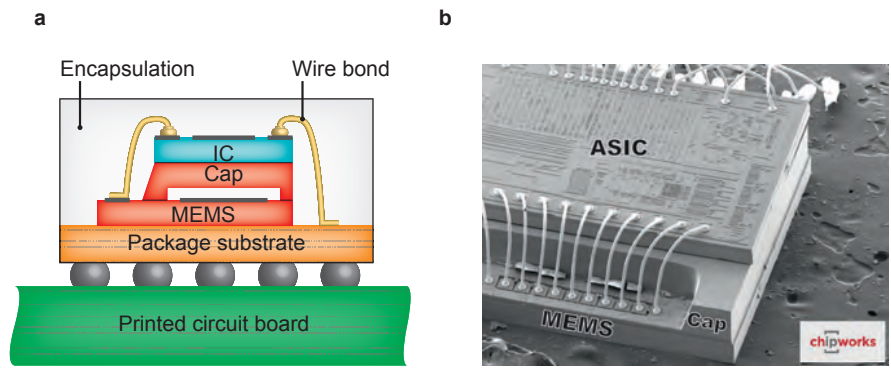

**Figure 3** (a) System-in-package solution constructed via 3D stacking with wire bonded interconnects. (b) SEM image of a decapsulated STMicroelectronics LIS331DLH 3-axis accelerometer (STMicroelectronics, Geneva, Switzerland). An ASIC chip is stacked on an encapsulated MEMS chip and interconnected via wire bonds. Package dimensions: $3 \times 3 \times 1$ mm$^3$. From Ref 54.

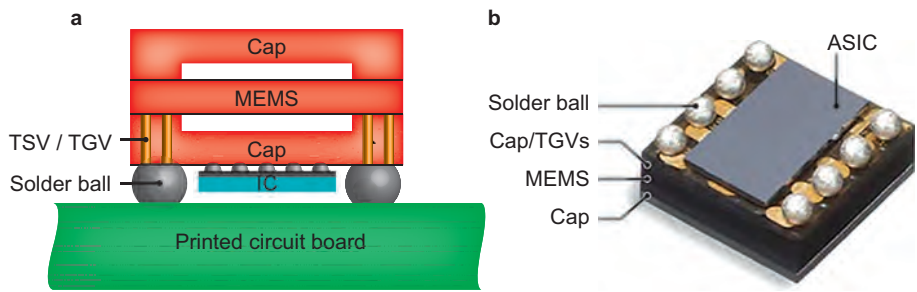

**Figure 4** (a) Chip-scale package (CSP): the MEMS and IC chips are attached via face-to-face flip-chip bonding. (b) Photograph of a 3-axis accelerometer (VTI, CMA 3000) fabricated using chip-on-MEMS technology. Package dimensions: $2 \times 2 \times 1$ mm$^3$. From Ref 61.

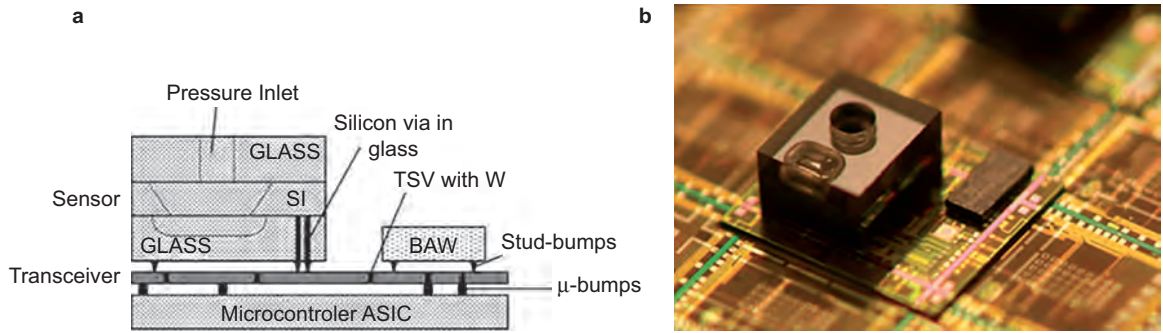

**Figure 5** (**a**) Sketch of a 3D sensor node stack fabricated using through-silicon vias. (**b**) Photograph of a 3D sensor node stack fabricated using through-silicon vias at the wafer level. From Ref 62.

common package[2]. As shown in Figure 5, complete sensor nodes comprising chip-level MEMS, ASIC, wireless communication and power management components are 3D integrated at the package level[62,63].

In summary, the key advantages of multi-chip solutions are their modularity, high flexibility and reasonably low fabrication complexity. In these approaches, the MEMS and IC manufacturing processes are completely decoupled, which results in full freedom of the design approaches and technology used for both the MEMS and IC chips. This enables rapid development cycles (short time-to-market) and relatively low development costs. The product specifications can easily be modified because the IC and MEMS chips are interchangeable. For the IC chips, the migration from one CMOS technology node to another, more advanced CMOS technology node can be readily implemented. Furthermore, the combined cost of a MEMS chip and the associated IC chip is not significantly impacted if there is a size difference between the MEMS and IC chips, as they are manufactured on separate wafers such that the maximum number of chips is fabricated on each wafer. Only tested (known-good) MEMS and IC chips are then packaged. Typically, the pre-packaging or sealing of MEMS structures is implemented at the wafer level, thus reducing the complexity of the final system. The manufacturing and testing infrastructure that is available from the semiconductor industry can be readily utilized for multi-chip solutions. Finally, technology, design and packaging standards can be established in a fairly simple manner for multi-chip solutions. The typical disadvantages of multi-chip solutions are their limited integration densities, large system footprints and thicknesses. Long electrical chip-to-chip connections can result in rather large parasitic capacitances and in the reduced robustness of electromagnetic compatibility (EMC), which can be a key disadvantage for certain devices, such as capacitive transducers. In addition, multi-chip solutions contain two or more chips per module, which can be a disadvantage in certain industries, such as the automotive industry, in which quality standards require the tracking of individual chips throughout the lifecycle of a product.

## 3 WAFER-LEVEL INTEGRATION OF MEMS AND ICS: SOC SOLUTIONS

The cofabrication of MEMS and IC components on a single substrate has been investigated for over three decades. Historically, however, few of those fabrication approaches have found their way into the industry. Only recently have various products implemented as SoC solutions, such as pressure sensors, inertial sensors, and microphones, been successfully commercialized.

SoC solutions are characterized by the fabrication and integration of MEMS and IC components on the same substrate, with chip separation occurring only at or near the end of the fabrication process. In general, SoC solutions can be categorized into two main integration schemes: (1) monolithic MEMS and IC integration techniques, as presented in Section 3.1, in which both the MEMS and IC structures are fabricated entirely on the same substrate and (2) heterogeneous MEMS and IC integration techniques, as presented in Section 3.2, in which the MEMS and IC structures are fully or partially prefabricated on separate substrates and subsequently merged onto a single substrate, typically via wafer bonding or similar transfer techniques.

### 3.1 SoC solutions using monolithic MEMS and IC integration

SoC solutions based on monolithic MEMS and IC integration can be categorized into four basic approaches: (1) monolithic MEMS and IC integration using MEMS-first processing, (2) monolithic MEMS and IC integration using interleaved MEMS and IC processing, (3) monolithic MEMS and IC integration using MEMS-last processing via the bulk micromachining of the IC substrate (also sometimes referred to as CMOS-MEMS) and (4) monolithic MEMS and IC integration using MEMS-last processing via layer deposition and surface micromachining.

#### 3.1.1 Monolithic MEMS and IC integration using MEMS-first processing

Monolithic MEMS and IC integration using MEMS-first processing is an integration approach in which all required processing steps for a complete MEMS device are performed prior to the CMOS processing, typically including substrate planarization to enable subsequent CMOS integration. This allows for a very high thermal budget of greater than 1100 °C for the processing of the MEMS material, which enables the use of high-temperature processes, to obtain high-performance epitaxial silicon or to release stress in thick layers of deposited poly-crystalline silicon layers, for example. Examples of MEMS-first processing schemes are illustrated in Figure 6 and 7.

Figure 6 illustrates the advanced porous silicon membrane process developed by Robert Bosch GmbH, Stuttgart, Germany, which is an innovative method for manufacturing vacuum cavities sealed with monocrystalline silicon membranes in silicon wafers[64–67]. In this process, localized porous silicon is created via anodic etching in hydrofluoric acid (HF). A subsequent annealing step in a hydrogen atmosphere at 900–1100 °C initiates a sintering process, causing the low-porosity layer near the surface to recrystallize and the high-porosity layer below to dissolve. An epitaxial silicon layer is then grown on top of the wafer, simultaneously creating a monocrystalline membrane and a high-quality silicon surface for CMOS fabrication. At present, Bosch is successfully fabricating pressure sensors for automotive applications using this method. Bosch's advanced porous silicon membrane process has been further developed using SOI substrates manufactured by SiTime Corp., Sunnyvale, CA, USA (now MegaChips, Japan) and Stanford University, CA, USA[68–70]. After the MEMS structures are patterned into the silicon device layer, the structures are encapsulated by an SiO$_2$ layer and a thin poly-silicon layer, as depicted in the schematic view in Figure 6d. Small vent holes, which allow for the release of the MEMS structures by means of HF vapor etching, are created in the poly-silicon layer. The vent holes are subsequently sealed by an additional epitaxial deposition of silicon. In areas outside the MEMS structures, the epitaxy process creates monocrystalline silicon, providing a high-quality material for CMOS fabrication. Figure 6e shows an SEM picture of a fully fabricated MEMS resonator structure. SiTime has successfully applied this MEMS-first process in high-volume fabrication of MEMS resonators for timing products[71].

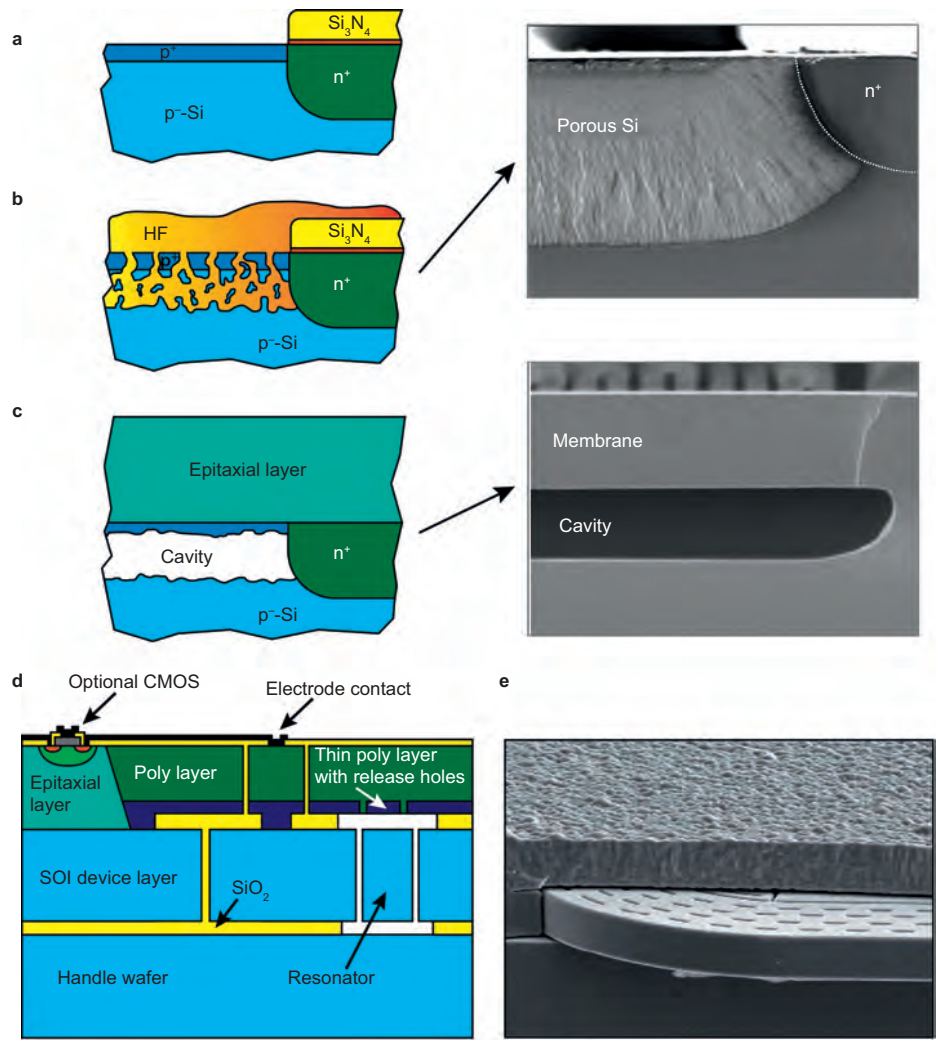

**Figure 6** (**a–c**) Advanced porous silicon membrane (APSM) process developed by Bosch. (**a**) n$^+$ and p$^+$ implantation. (**b**) Anodic HF etching. (**c**) Sintering to form a cavity and a silicon epitaxial layer. Adapted from Ref 65 and Ref 66. (**d–e**) Similar MEMS-first platform developed by SiTime (now MegaChips, Japan) and Stanford University. MEMS structures are fabricated in the SOI device layer and released by means of HF vapor etching. The ICs are fabricated in the epitaxially grown silicon top layer. (**d**) Schematic cross-section. (**e**) SEM image of a fabricated resonator[72].

Another MEMS-first platform is the 'plug-up' process developed by VTT, which has found commercial application in pressure sensors[73–75]. As shown in Figure 7, this process is based on an SOI substrate in which cavities are formed via HF etching of the buried oxide through small holes in the silicon device layer that are covered by an HF-permeable layer of poly-silicon. The holes are then filled with another layer of poly-silicon, and the surface is planarized via chemical-mechanical planarization (CMP) to expose the monocrystalline surface, onto which the CMOS circuits can then be integrated.

In summary, MEMS-first integration approaches offer favourable conditions for MEMS fabrication, such as a very high thermal budget. This allows for the fabrication of high-performance MEMS structures and hermetically sealed packages of superior quality, which is why this technology is used for products such as high-performance MEMS resonators. However, there are strict requirements on the surface planarity and material exposure of the pre-processed MEMS wafers. Typically, it is not permitted for pre-processed wafers to be brought into standard CMOS fabs, which is why this approach is utilized predominantly by companies that have access to a dedicated CMOS fab. For the same reason, this approach is impractical in fabless business models.

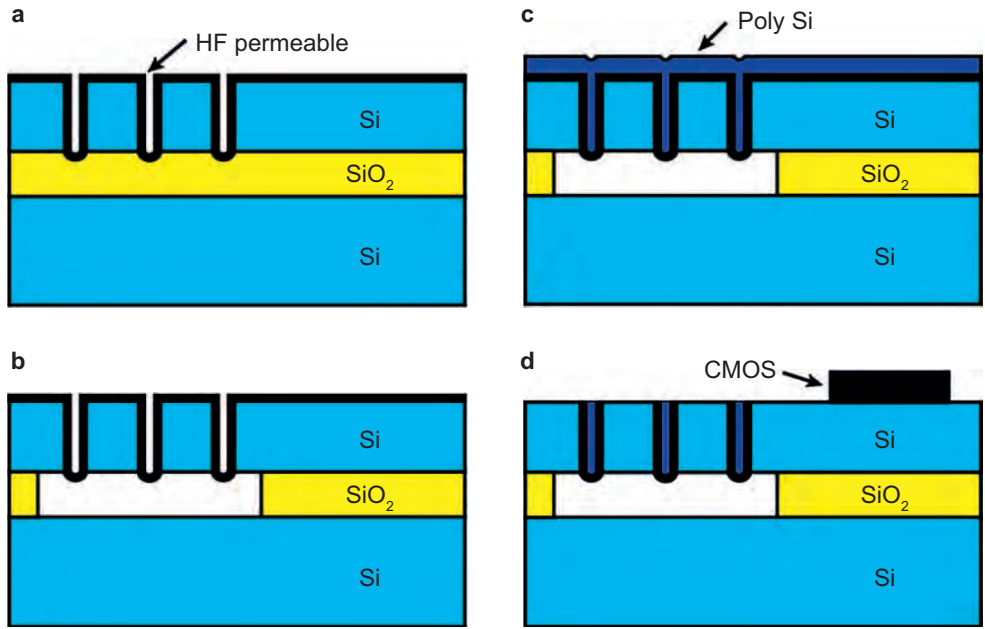

**Figure 7** Processing sequence for the 'plug-up' concept. (**a**) Small holes are etched, and a thin HF-permeable poly-silicon layer is deposited. (**b**) The buried oxide is etched with HF. (**c**) The holes are closed via the deposition of poly-silicon. (**d**) CMP followed by CMOS processing. Adapted from Ref 73.

3.1.2 Monolithic MEMS and IC integration using interleaved MEMS and IC processing

Monolithic MEMS and IC integration using interleaved MEMS and IC processing is achieved through a combination of MEMS processing steps performed before, after, or during CMOS fabrication. Examples of interleaved MEMS and IC processing platforms include $M^3$EMS, Mod-MEMS and various SOI-based integration concepts.

The $M^3$EMS technology platform[76], developed by Sandia National Laboratories, California, Albuquerque, NM, USA, utilizes shallow trenches in which MEMS structures are fabricated through a series of silicon oxide and poly-silicon deposition steps. A subsequent CMP process is applied to planarize the wafer and expose the monocrystalline silicon surface for IC fabrication.

A post-CMOS HF etching step is necessary to release the MEMS structures and to complete device fabrication. In Figure 8, a schematic cross-section of a fully fabricated device is presented.

The Mod-MEMS integration process, developed by Analog Devices Inc., Norwood, MA, USA[77,78] in collaboration with UC Berkeley, uses a different technique to achieve the planarity necessary for CMOS fabrication. As illustrated in Figure 9a, the MEMS structures are manufactured on top of a flat silicon wafer via the deposition and surface micromachining of silicon nitride, silicon oxide and poly-silicon. Monocrystalline silicon is epitaxially grown on the area beside the MEMS structures to level out the surface topography. Finally, the surface is polished via CMP and prepared for subsequent CMOS processing. The fabrication is completed with an HF-based release etch.

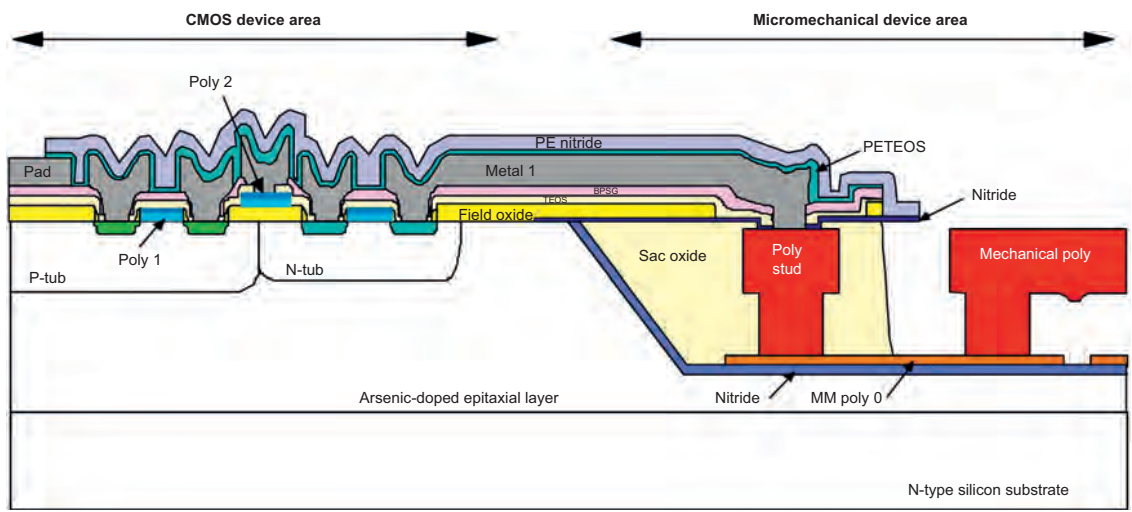

**Figure 8** Multilevel MEMS formed within a shallow trench using the Sandia National Laboratories $M^3$EMS platform. From Ref 76.

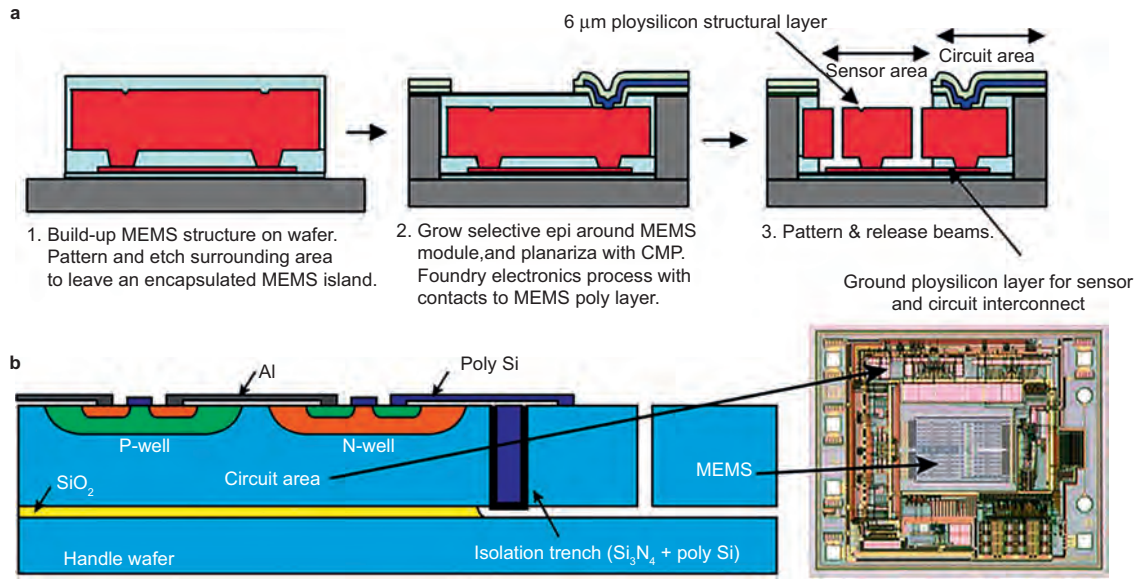

**Figure 9** (**a**) Process outline for the Mod-MEMS platform. From Ref 77. (**b**) Schematic cross section and photograph of an SOI-MEMS device, showing the released MEMS structure fabricated from an SOI wafer, the isolation trench, and the IC area. Adapted from Ref 6 and Ref 80.

A related SOI-based integration technology that was also developed by Analog Devices Inc., Norwood, MA, USA[79] in collaboration with UC Berkeley[80–82] is depicted in Figure 9b. This integration platform involves separating the MEMS area from the CMOS area by creating isolation trenches in the silicon device layer of an SOI wafer. After a CMP step, the CMOS is fabricated in the device layer, followed by the definition of the MEMS structure and its release etching.

The CNM Institute of Microelectronics in Barcelona, Spain, has also developed an SOI-based integration platform in which the MEMS structures are fabricated in the device layer, whereas the CMOS circuits are fabricated in the bulk silicon[83]. This optical iMEMS technology platform[84] was developed for the integration of high-quality micro-mirrors with a BiCMOS process. Instead of using standard SOI wafers, the iMEMS process is based on a three-layer silicon stack that is formed through wafer bonding. A similar technology platform has also been developed at MIT[85]. MEMS interleaved processing can also be achieved by introducing additional, custom processing steps into the regular CMOS processing sequence[86]. This approach has found commercial application in the fabrication of pressure sensors by Infineon Technologies AG, Neubiberg, Germany[87] and inertial sensors by Analog Devices Inc., Norwood, MA, USA[88,89]. The University of Michigan has introduced a method based on a $p^{++}$ etch-stop technique. Here, an additional implantation step in the CMOS process defines highly doped regions, and the MEMS structures are subsequently created via the bulk machining of the substrate. A large variety of applications using this technique have been demonstrated, ranging from pressure sensors and infrared sensors to neural probes[90–95].

In summary, monolithic MEMS and IC integration using interleaved MEMS and IC processing offers the possibility of integrating high-performance MEMS materials and devices together with CMOS circuits on the same substrate. In most approaches, the MEMS and CMOS structures are placed side-by-side on the substrate, which limits the achievable integration density to some extent. MEMS interleaved approaches require full access to a dedicated custom CMOS production line, which considerably limits the general utility of the technology, as this requirement is typically not compatible with fabless business models.

### 3.1.3 Monolithic MEMS and IC integration using MEMS-last processing via the bulk micromachining of the IC substrate

Monolithic MEMS and IC integration using MEMS-last processing by means of bulk micromachining is achieved through integration schemes in which the MEMS structures are manufactured after the entire CMOS fabrication process has been completed. This includes all approaches in which the MEMS components are fabricated in the CMOS substrate and/or in the CMOS thin-film layers; approaches that include subsequent layer deposition are presented in Section 3.1.4. MEMS-last integration via the bulk micromachining of the IC substrate has been under investigation since the late 1970s. Early efforts consisted of backside micromachining to realize piezoresistive pressure sensors[96,97]. Since that time, a variety of different fabrication approaches have been developed, as shown in Figure 10.

Backside etching with timed etch-stop control, as illustrated in Figure 10b, has been commercialized by Freescale for pressure sensor applications[99]. An alternative to timed etch-stop control is the electrochemical etch-stop technique depicted in Figure 10c, which allows for very precise etch stops between p-doped and $n^+$-doped regions[100,101]. Various MEMS structures, such as membranes, cantilevers and suspended islands, can be fabricated using this technique. Another approach is to directly structure the front side of the substrate. The technology platform shown in Figure 10d was developed at Carnegie Mellon University and uses the back-end interconnect metal layers as a hard mask for the patterning of the MEMS structures[102]. This method has found application in the fabrication of a variety of devices, ranging from accelerometers[103] and gyroscopes[104] to infrared imagers[105] and variable capacitors[106]. A similar fabrication approach, which uses photoresist masks for MEMS structuring, is depicted in Figure 10e. This method has been applied for the fabrication of passive, free-etched inductors for RF circuits[107] and, more recently, for micro-bolometers[108] as well as single-chip scanning microwave microscope and atomic force microscope (AFM) systems[109]. The method presented in Figure 10f is based on the deep-reactive ion etching (DRIE) of both the front and back sides of the substrate. This enables the fabrication of integrated monocrystalline silicon MEMS structures with superior mechanical properties[110]. The fabrication of infrared bolometer arrays using an electrochemical etch stop has been developed at the Middle East Technical University, Ankara, Turkey[111], and is being commercialized by MikroSens, Ankara, Turkey[112]. In this process, n-well micro-bolometers are created via a standard CMOS process and connected by dielectric support arms, with the metal interconnect layer providing the electrical connections. A front-side electrochemical etch-stop technique is used to free etch the bolometer pixels without etching away the n-wells.

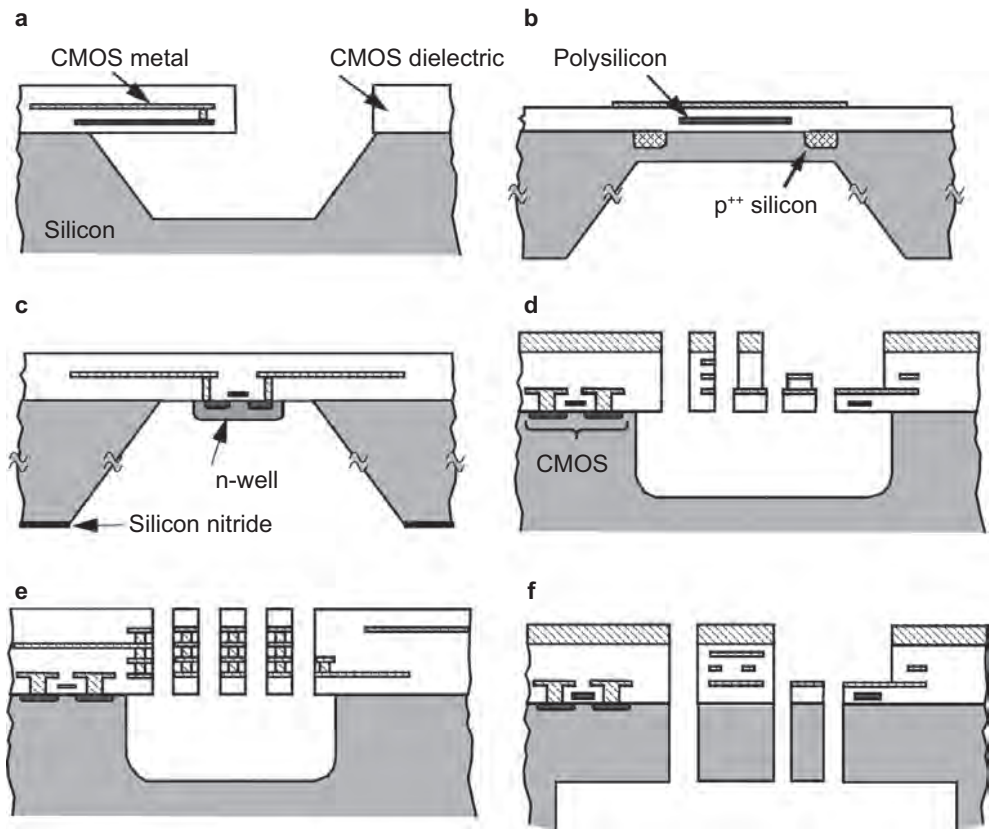

**Figure 10** Various approaches to monolithic MEMS and IC integration using MEMS-last processing via the bulk micromachining of the IC substrate: (**a**) Front-side wet etching. (**b**) Backside wet etching. (**c**) Backside wet etching using an electrochemical etch stop. (**d**) Front-side dry etching using a metal hard CMOS mask. (**e**) Front-side dry etching using a photoresist mask. (**f**) Front-and backside anisotropic deep-reactive ion etching (DRIE) of a CMOS wafer. Adapted from Ref 98.

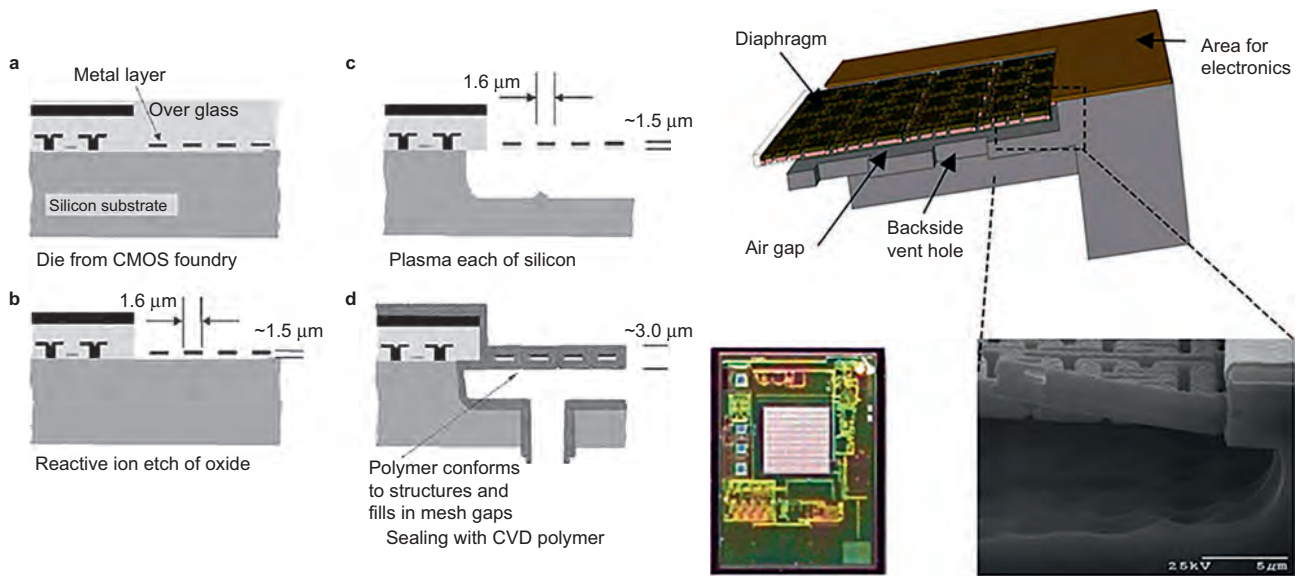

**Figure 11** Fabrication process and images of an IC-integrated MEMS microphone. Adapted from Ref 6 and Ref 113.

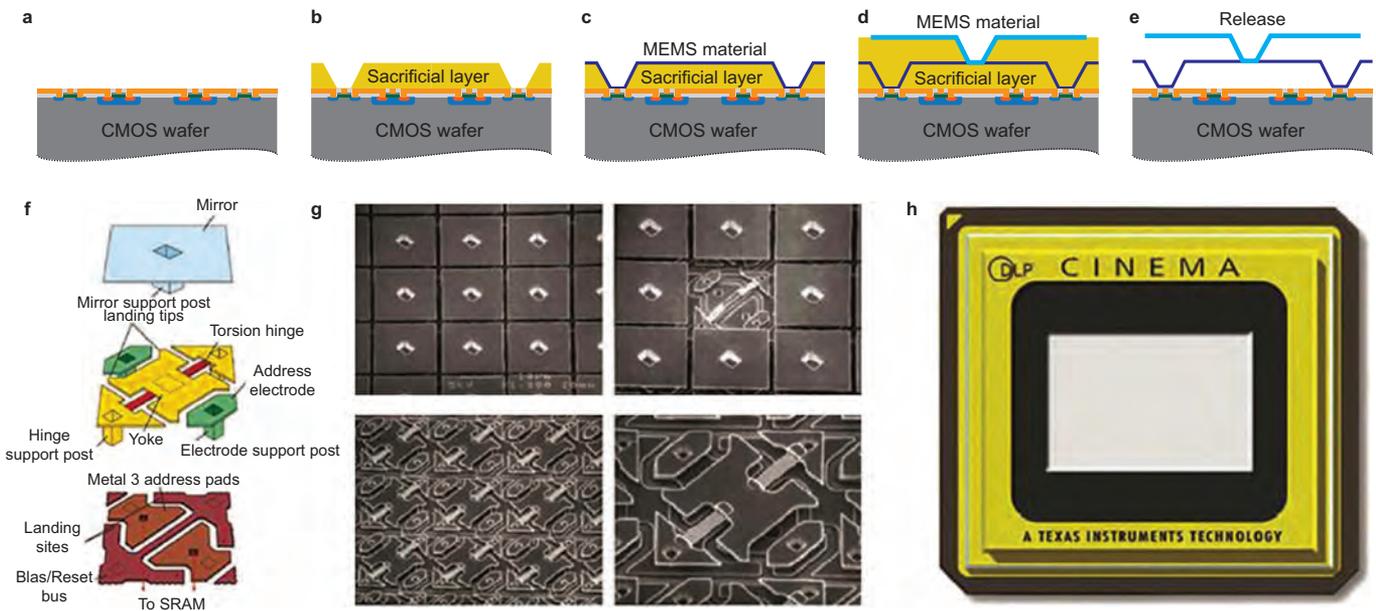

**Figure 12** Typical processing scheme for monolithic MEMS-on-IC integration using MEMS-last processing via the deposition and surface micromachining of two layers: (**a**) Manufacturing of the CMOS wafer. (**b**) Deposition and patterning of the sacrificial layer. (**c–d**) Deposition and patterning of the first layer of MEMS material, the second sacrificial layer and the second layer of MEMS material. (**e**) Etching of the sacrificial layer. (**f**) Schematic diagram of a deflected torsional micro-mirror pixel. (**g**) SEM images of micro-mirror pixels with the mirror plate of the center pixel removed to expose the underlying hinge structures. (**h**) Packaged DLP$^{TM}$ chip containing an array of more than two million micro-mirrors, placed in a ceramic package with a glass window[6].

An integration process for MEMS microphones based on a combination of DRIE backside etching and the use of the interconnect metal as a hard mask for front-side etching was developed at Carnegie Mellon University and later commercialized by Akustica, Inc. (now Bosch)[113]. The process utilizes a serpentine mesh pattern in a metal interconnect layer to form a suspended membrane that is sealed by a CVD-deposited polymer. Figure 11 shows a schematic illustration of the fabrication process for the microphone membrane as well as images of a fabricated device.

A variety of different MEMS-last integration approaches based on bulk micromachining have been applied in the fabrication of monolithic inkjet printheads with integrated CMOS circuits[114–116]. Inkjet printheads represent the largest-volume commercial application of monolithic MEMS and CMOS integration in the industry to date.

In summary, monolithic MEMS and IC integration using MEMS-last processing by means of bulk micromachining has the important advantage that it can be implemented using existing IC infrastructure, including access to design tools and IC foundries. The MEMS components are formed in the completed IC wafers using fairly simple and cost-effective processing steps. This approach offers the potential for short development cycles (i.e., short times to market) and the use of extremely low-cost components. The disadvantages of this approach are the limited design freedom for the MEMS devices and the limited selection of materials for the MEMS devices allowed by the CMOS process. The etching processes, such as the passivation step, must be carefully designed not to attack any CMOS structures. In addition, in standard CMOS production lines, the mechanical properties of the materials and layers are typically not very well characterized or controlled. This can result in reliability and repeatability problems in the fabricated MEMS devices.

### 3.1.4 Monolithic MEMS and IC integration using MEMS-last processing via layer deposition and surface micromachining

In monolithic MEMS and IC integration using MEMS-last processing via layer deposition and surface micromachining, the MEMS structures are fabricated by depositing and micromachining materials on top of completed CMOS wafers. Figure 12 illustrates a characteristic processing scheme for the integration of two-layer MEMS structures on top of a CMOS substrate. First, a sacrificial layer is deposited and patterned on the CMOS substrate, as shown in Figure 12a and b. Then, the MEMS material is deposited and patterned, as shown in Figure 12c. Figure 12d illustrates a second cycle of sacrificial material deposition and patterning followed by MEMS material deposition and patterning to create the second layer of the MEMS structure. Finally, the sacrificial material is removed in a release etching step to yield a suspended MEMS device on top of the CMOS substrate, as depicted in Figure 12e. This approach allows for the realization of rather complex MEMS structures on top of IC substrates[6,117]. In this processing scheme, it is important that the procedures for the deposition of the sacrificial and MEMS materials are compatible with the CMOS substrate. The maximum temperature to which a typical CMOS substrate can be exposed without causing permanent damage to the circuits is approximately 400–450 °C[6]. Some of the most commonly used sacrificial materials are various types of polymers that can be easily deposited and then removed using oxygen plasma dry-etching processes with very high etching selectivity towards the typical materials that are used for MEMS and CMOS structures[6,117,118]. Metals[6], dielectrics[119] and poly-crystalline and amorphous semiconductors[6,120] have also been used as sacrificial layers. The functional MEMS materials that have been deposited on CMOS substrates include various metals[6,118,119,121], silicon nitride, silicon dioxide[6,122], piezoelectric aluminium nitride[6], zinc-oxide[6], poly-crystalline silicon germanium[120,123,124], amorphous silicon[117,122], vanadium oxide[122,125], structural polymers[6,126] and various combinations of the above materials.

The vast majority of commercially available micro-mirror arrays[118] and uncooled infrared bolometer focal plane arrays[122] are implemented as SoC solutions based on monolithic MEMS and IC integration using MEMS-last processing via layer deposition and surface micromachining. One example are the digital micro-mirror arrays from Texas Instruments that are depicted in Figure 12f–h[6,118].

The application of monolithic MEMS and IC integration using MEMS-last processing via layer deposition and surface micromachining has been proposed for a large number of MEMS devices, including infrared bolometer arrays consisting of $VO_x$[125], amorphous silicon[117] or metal[121], micro-mirror arrays consisting of poly-crystalline silicon germanium[123] or metal[119], inertial sensors consisting of poly-crystalline silicon germanium[124], bio-chips consisting of structural polymers[127], bulk acoustic wave resonators consisting of poly-crystalline silicon germanium[128], and many other MEMS devices based on a variety of materials[6].

In summary, the most important advantage of monolithic MEMS and IC integration using MEMS-last processing via layer deposition and surface micromachining is that in this approach, standard CMOS foundries can be utilized to fabricate the CMOS wafers. The MEMS structures are integrated on top of completed CMOS wafers, utilizing, e.g., a specialized MEMS fab. The complexity of the process flow is relatively low, and the same chip area can be simultaneously occupied by both MEMS and CMOS structures. Thus, the available area on the chip is used very efficiently, and the achievable integration densities can be extremely high. The primary disadvantage of this approach is that the deposition temperature for the MEMS materials must lie within the allowed temperature budget for the CMOS wafer, which is typically below 400 or 450 °C. This excludes the use of important high-performance MEMS materials such as mono-crystalline and poly-crystalline silicon and makes the technology less attractive for MEMS devices that must be implemented using high-performance MEMS materials, such as inertial sensors and resonators. Such devices typically benefit from the outstanding material properties provided by monocrystalline silicon, including low internal stresses, perfect elastic behaviour and a high quality factor (Q).

### 3.2 SoC solutions using heterogeneous MEMS and IC integration

Heterogeneous MEMS and IC integration is defined here as the joining of two or more substrates that contain fully or partially fabricated MEMS and IC structures to produce a heterogeneous SoC solution. Heterogeneous MEMS and IC integration can be categorized as either (1) heterogeneous MEMS and IC integration with via formation during layer transfer (e.g., wafer bonding), also referred to as via-first processes, or (2) heterogeneous MEMS and IC integration with via formation after layer transfer (e.g., wafer bonding), also referred to as via-last processes.

#### 3.2.1 Heterogeneous MEMS and IC integration with via formation during layer transfer

An example of a typical heterogeneous MEMS and IC integration process in which the vias that establish the mechanical and electrical contacts between components on the two substrates are formed during the layer transfer process (i.e., wafer bonding) is shown in Figure 13a. This specific integration process is used for the manufacturing of very high-volume multi-axis gyroscope products distributed by InvenSense in the USA (manufactured by TSMC in Taiwan)[9,11,129]. In this process, the monocrystalline silicon MEMS sensors (e.g., gyroscopes), including the cap for the MEMS package, are prepared on an initial wafer that is subsequently bonded to a pre-fabricated CMOS-based IC wafer that contains etched cavities, as indicated in Figure 13a–c. In this example, aluminium/germanium eutectic bonding is used to bond and seal the combined gyro and cap wafer against the aluminium metal layer on the CMOS wafer[130]. Figure 13d shows SEM images of a gyroscope that was manufactured using this technology[9].

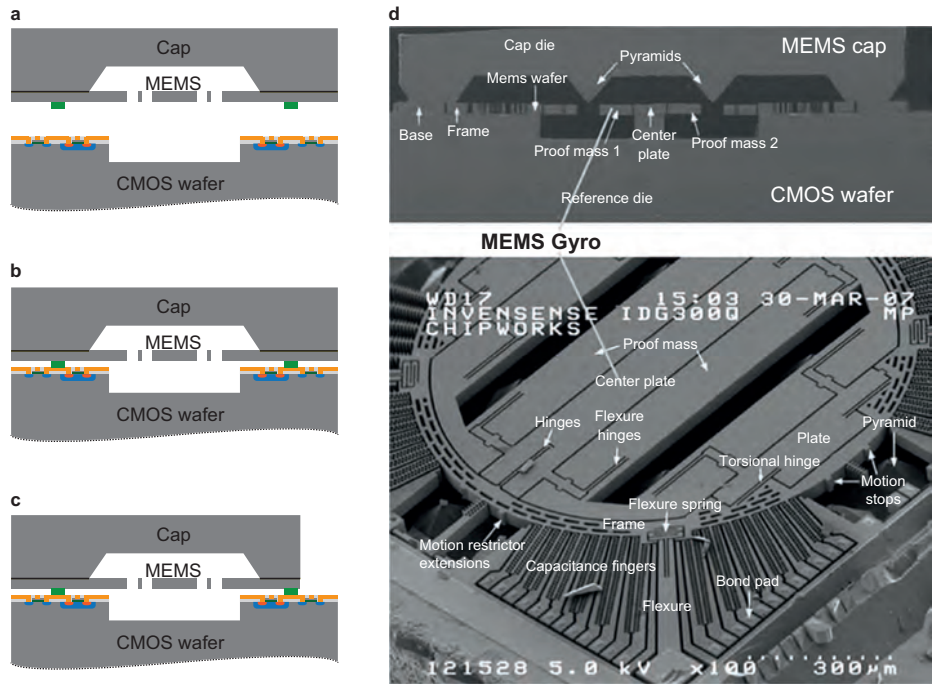

**Figure 13** Via-first heterogeneous integration platform from InvenSense Inc., San Jose, CA, USA, which is used for, e.g. the high-volume manufacturing of gyroscope products: (**a**) A pre-fabricated monocrystalline silicon MEMS sensor, including a cap, is (**b**) bonded to a pre-fabricated CMOS-based IC wafer containing etched cavities; then, (**c**) the contact pads on the CMOS wafer are revealed. (**d**) SEM images of a gyroscope integrated with CMOS circuits. From Ref 9.

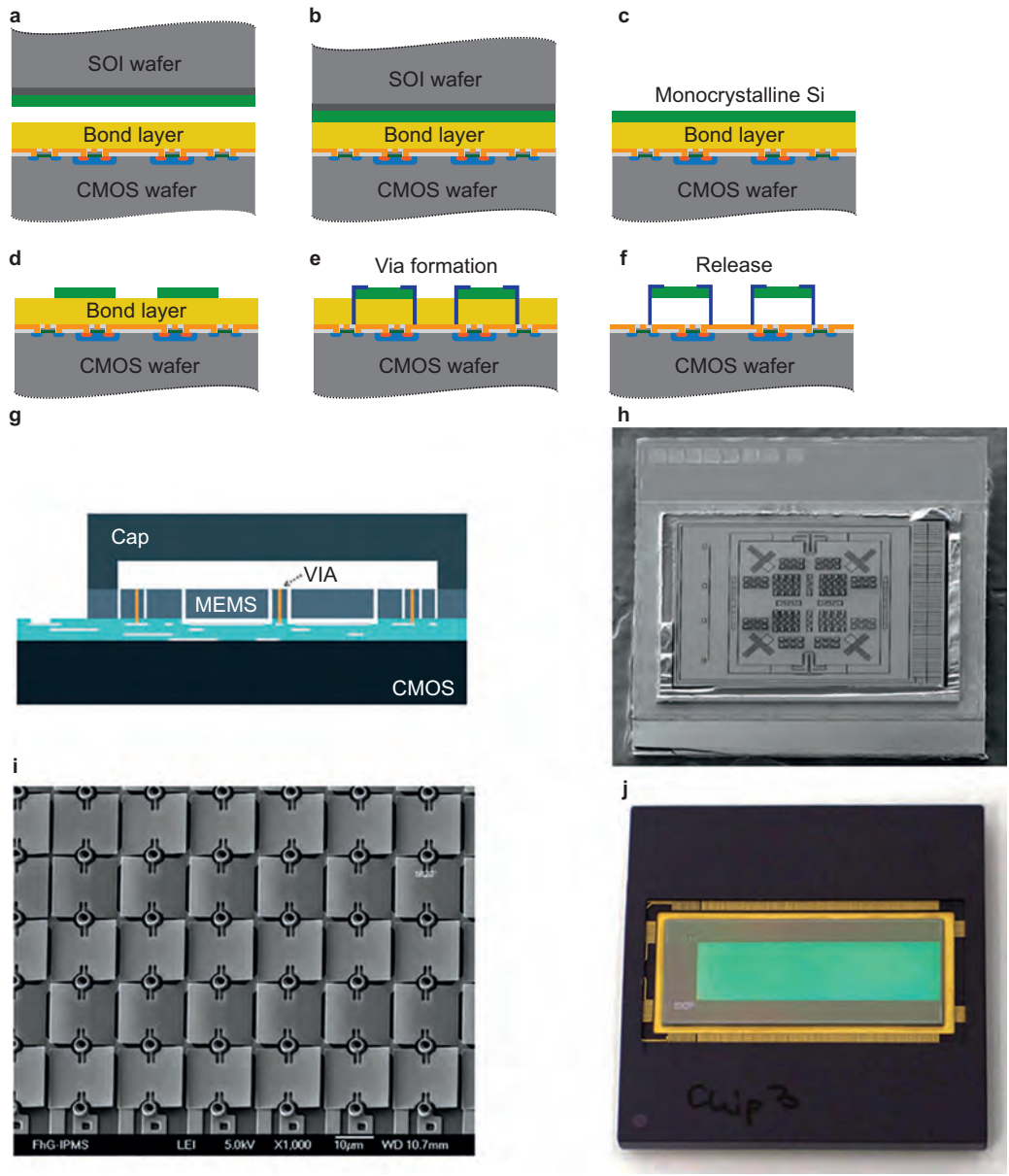

**Figure 14** Example of heterogeneous MEMS and IC integration with via formation after bonding: (**a**) Individual fabrication of the IC wafer and the handle wafer, which contains a monocrystalline silicon MEMS device layer (e.g., an SOI wafer). (**b**) Wafer bonding using an intermediate adhesive layer. (**c**) Sacrificial dissolution or release of the handle wafer. (**d**) Patterning of the monocrystalline silicon. (**e**) Via-hole etching and via formation. (**f**) Sacrificial etching of the intermediate adhesive layer. (**g**) Design based on via-last heterogeneous MEMS and IC integration for accelerometers distributed by mCube in the USA (manufactured by TSMC Ltd.) and (**h**) SEM image of an mCube accelerometer. From Ref 142. SEM images of (**i**) a 1-megapixel monocrystalline silicon mirror array on CMOS driving electronics and (**j**) a packaged micro-mirror array. From Ref 143.

One of the first reported examples of heterogeneous MEMS and IC integration with via formation during wafer bonding was a monocrystalline silicon mirror array consisting of 32 × 32 mirrors with individual mirror dimensions of 1 mm × 1 mm[131,132]. Another well-developed heterogeneous MEMS and IC integration platform has been used to manufacture large arrays of IC-integrated AFM tips[133] and RF switches[134,135]. In this process, a glass wafer with pre-fabricated monocrystalline silicon AFM tips is aligned with and bonded to a CMOS-based IC wafer using a combination of solder-bump bonding and polyimide adhesive bonding. Next, the transferred AFM tips are released from the glass wafer via laser debonding. The vias of the AFM tips are 15 µm in diameter, the AFM tip array has a pitch of 130 µm × 100 µm, and the monocrystalline silicon AFM cantilevers are 300 nm thick. Other variations of heterogeneous MEMS and IC integration with via formation during device transfer have been proposed[22,25], including the integration of carbon-based NEMS materials[136–138], material integration using self-assembly processes[139], and transfer printing techniques for MEMS[140].

In summary, a key advantage of heterogeneous MEMS and IC integration with via formation during layer transfer is that it allows high-performance MEMS materials such as monocrystalline silicon to be combined with standard CMOS-based IC wafers. The MEMS components can be pre-fabricated prior to integration with the IC wafer, and the integration and packaging can be performed in a single bonding step, as illustrated in Figure 13. In addition, some of these technologies are supported by the existing foundry infrastructure, thus greatly facilitating fabless business models. The disadvantages of via-first processes are that they often require aligned substrate-to-substrate bonding, which adds process complexity and results in limitations on the achievable post-bonding alignment accuracies[141]. Additionally, reliable electrical interconnections of bonded vias with dimensions of less than 10 μm are challenging to implement.

### 3.2.2 Heterogeneous MEMS and IC integration with via formation after layer transfer

Heterogeneous integration with via formation after layer transfer (i.e., wafer bonding) is schematically illustrated in Figure 14. In this approach, the vias that establish the mechanical and electrical contacts between components on the different substrates are defined after the layer transfer is performed, as depicted in Figure 14e. In this specific example, the MEMS handle substrate is sacrificially dissolved or released from the MEMS structures prior to via formation[9].

A heterogeneous MEMS and IC integration process with via formation after wafer bonding has been developed by TSMC, Ltd. in Taiwan and is offered as a MEMS foundry platform[11,144]. This process involves bonding a completed CMOS wafer onto a silicon MEMS wafer using an intermediate $SiO_2$ adhesive layer and subsequently forming vias to electrically interconnect the MEMS devices to the CMOS circuits. This process is utilized for, e.g., the high-volume manufacturing of multi-axis accelerometers by mCube Inc., San Jose, CA, USA [see Figure 14g–h][145].

Different heterogeneous MEMS and IC integration processes with via formation after layer transfer have also been proposed for the realization of infrared bolometer arrays[122,146–152], arrays of tilting and piston-type micro-mirrors[143,153–156], nanoelectromechanical switches for logic circuits[157], micro-Pirani vacuum gauges[158] and RF MEMS devices[159–163]. In most of these integration platforms, an intermediate polymer adhesive layer is used in the wafer bonding process. The advantages of adhesive bonding using an intermediate polymer layer are the high process yield that can thus be achieved and the fact that no demanding surface pre-treatment or surface planarization steps are required for the bonding[164–166]. The first reported MEMS device to be integrated on top of functional CMOS circuits using via-last heterogeneous 3D integration was a 1-megapixel monocrystalline silicon micro-mirror array, depicted in Figure 14i–j[143]. This mirror array has a pixel pitch of 16 μm × 16 μm and contains silicon mirror membranes that are 340 nm thick and are located at an extremely well-defined distance of 700 nm from the corresponding electrodes on the underlying CMOS circuits. The mirror vias have a diameter of 2 μm, and the torsional mirror hinges are 600 nm wide. Heterogeneous MEMS and IC integration with via formation after layer transfer has also been proposed for the integration of carbon-based NEMS materials[167], self-assembly processes[139] and transfer printing techniques[140].

In summary, the key advantage of heterogeneous MEMS and IC integration processes with via formation after layer transfer is the fact that they allow high-performance MEMS materials such as monocrystalline silicon to be combined with standard CMOS-based IC wafers. Therefore, wafer-to-wafer alignment during bonding is not necessarily required, and as a consequence, the achievable placement accuracy of the MEMS components on the IC wafer is defined by the placement accuracy that can be achieved using the lithography tools, which can easily be in the nm-range. The feasible via dimensions can be in the sub-μm-range, and the distance between the MEMS device membranes and the IC substrate surface can be accurately defined by the thickness of the intermediate bonding layer over a large interval of below 100 nm to several tens of μm. Thus, via-last heterogeneous MEMS and IC integration platforms can be employed for the fabrication of IC-integrated MEMS with extremely small dimensions and extremely high integration densities. A disadvantage of this approach is the greater number of processing steps as compared to typical via-first heterogeneous MEMS and IC integration processes.

## 4 OUTLOOK AND CONCLUSIONS

A wide variety of alternative solutions for combining and integrating MEMS and ICs are available. These methods can be divided into hybrid multi-chip solutions and SoC solutions, both of which have distinct advantages and disadvantages. In general, multi-chip solutions are more flexible and less complex, allow for rapid product development cycles and are cost-effective for all combinations of MEMS and IC chip sizes. However, compared with SoC solutions, multi-chip solutions offer relatively low integration densities, are larger in size and footprint and suffer from lower EMC robustness and high parasitic capacitances in the electrical connections between the MEMS and IC components. SoC solutions, on the other hand, have the disadvantages of higher complexity, lower flexibility and longer development times. For certain classes of MEMS products, such as large arrays of transducers, the high integration density that can be achieved in SoC solutions is a necessary prerequisite for the implementation of these products. High integration densities also endow MEMS products with small system dimensions, including a small footprint and low thickness.

If the fabrication yield of either the IC or the MEMS process is low, then SoC solutions are strongly affected by the aggregated combined yield, whereas in the multi-chip approach, known good dies can be paired to achieve a better overall yield. Another cost-driving factor for SoC solutions arises if the MEMS and IC chips are significantly different in size. This leads to unused wafer area, which, in many cases, is not economically viable.

The potentially lower yield of SoC solutions is offset by their lower costs for testing and packaging. SoC solutions offer the possibility of packaging at the wafer level, which reduces the number of packaging and wire bonding steps that must be performed at the chip level. Testing is also often a significant contribution to the overall system cost. For SoC solutions, this cost can be greatly reduced because only one test on the final assembled device is needed, compared with the several tests of individual modules prior to assembly as well as tests after assembly that are required in the case of multi-chip solutions.

The development of cost-effective through-substrate vias of increasingly smaller pitch and thinner chips has had a profound impact on MEMS and IC integration based on hybrid multi-chip solutions. The 3D stacking of IC and MEMS chips with small through-substrate vias enables the fabrication of relatively thin devices with small footprints and electrical interconnects with small parasitic capacitances, while still maintaining the advantages of hybrid multi-chip solutions. Extensive research efforts by the semiconductor industry in the areas of 3D IC integration and through-substrate vias will facilitate the widespread adoption of these solutions for integrating MEMS and ICs, as well.

Heterogeneous MEMS and IC integration approaches for SoC solutions enable the integration of high-performance MEMS materials and devices on top of standard foundry CMOS wafers. Thus, no compromise must be made in the selection of MEMS materials, and the most suitable CMOS technology can be chosen. The same chip area can be simultaneously occupied by both MEMS and CMOS structures. These advantages, together with the development of stable and CMOS-compatible wafer-level packaging solutions, enable flexible and cost-effective SoC solutions. Several products that are based on heterogeneous MEMS and IC integration approaches have already gained significant market shares on the extremely competitive high-volume consumer market[9,11,12]. Some of the factors contributing to this success are the compatibility of such approaches with fabless and fab-light business models, their ability to utilize standard CMOS-based ICs from various foundry sources and the potential they offer to dramatically shrink the overall MEMS device dimensions, thereby enabling the combination of multiple sensors on a single, highly integrated chip.

In summary, future developments in multi-chip and SoC solutions are converging towards higher MEMS and IC integration densities and, thus, smaller and cheaper components. A clear trend that is evident in MEMS sensor components is the integration of several sensing functions (e.g., multi-axis inertial and magnetic field sensing) in a single module, together with specialized processing functions for the sensor signals. These pre-processed and robust (often digital) sensor signals can be interfaced with high-performance processing units for advanced signal processing and the fusion of signals from various sensor elements.

## ACKNOWLEDGEMENTS
This review paper was inspired by a two-day course with the title "Cofabrication of MEMS and Electronics" that was organized in November 2008 by IMEC in Leuven, Belgium. The authors thank the organizer, lecturers and participants of this course. The work was partially funded by the Swedish Research Council, by the European 7th Framework Programme under grant agreement FP7-NEMIAC (No. 288670) and by the European Research Council through the ERC Advanced Grant xMEMs (No. 267528) and the ERC Starting Grant M&M's (No. 277879).

## COMPETING INTERESTS
The authors declare no conflict of interest.

## REFERENCES
1   ITRS. International Technology Roadmap for Semiconductors, 2011 Edition. The International Technology Roadmap, 2011; http://www.itrs.net/reports.html.
2   Tummala RR. SOP: What is it and why? A new microsystem-integration technology paradigm-Moore's law for system integration of miniaturized convergent systems of the next decade. *IEEE Transactions on Advanced Packaging* 2004; **27**: 241–249.
3   Davis WR, Wilson J, Mick S et al. Demystifying 3D ICs: The pros and cons of going vertical. *IEEE Design & Test of Computers* 2005; **22**: 498–510.
4   Yole Développement. Status of the MEMS Industry in 2014. Yole Développement, Villeurbanne, France, 2014; http://www.i-micronews.com/mems-sensors-report/product/status-of-the-mems-industry-2014.html.
5   Jérémie Bouchaud. MEMS Market Tracker – Q4 2014. IHS Technology, Englewood, CO 80112, USA, 2015; https://technology.ihs.com/490495/mems-market-tracker-q4-2014.
6   Fedder GK, Howe RT, Liu T-JK, Quevy EP. Technologies for cofabricating MEMS and electronics. *Proceedings of the IEEE* 2008; **96**: 306–322.
7   Brand O. Microsensor integration into systems-on-chip. *Proceedings of the IEEE* 2006; **94**: 1160–1176.
8   French PJ, Sarro PM. Integrated MEMS: Opportunities and challenges. In: Kahrizi M (ed). Micromachining Techniques for Fabrication of Micro and Nano Structures. InTech Publishing, Rijeka, Croatia, 2012. DOI: 10.5772/1364.
9   Lapisa M, Stemme G, Niklaus F. Wafer-level heterogeneous integration for MOEMS, MEMS, and NEMS. *IEEE Journal of Selected Topics in Quantum Electronics* 2011; **17**: 629–644.
10  Esashi M. Wafer level packaging of MEMS. *IOP Journal of Micromechanics and Microengineering* 2008; **18**: 073001.
11  Niklaus F, Lapisa M, Bleiker SJ et al. Wafer-level heterogeneous 3D integration for MEMS and NEMS. 2012 3rd IEEE International Workshop on Low Temperature Bonding for 3D Integration (LTB-3D); 22–23 May 2012; Tokyo, Japan; 2012: 247–252.
12  Esashi M, Tanaka S. Heterogeneous integration by adhesive bonding. *Micro and Nano Systems Letters* 2013; **1**: 1–10.
13  O'Neal CB, Malshe AP, Singh SB et al. Challenges in the packaging of MEMS. International Symposium on Advanced Packaging Materials: Processes, Properties and Interfaces; 14–17 Mar 1999; Georgia, USA; 1999: 41–47.
14  Malshe AP, O'Neal C, Singh SB et al. Challenges in the packaging of MEMS. *International Journal of Microcircuits and Electronic Packaging* 1999; **22**(3): 233–241.
15  Mirza AR. Wafer-level packaging technology for MEMS. The Seventh Intersociety Conference on Thermal and Thermomechanical Phenomena in Electronic Systems; 23–26 May 2000; Las Vegas, NV, USA; 2000: 119.
16  Esashi M. Wafer level packaging of MEMS. The 15th International Conference on Solid-State Sensors, Actuators and Microsystems (TRANSDUCERS'2009), 21–25 Jun 2009; Denver, CO, USA; 2009: 9–16.
17  Hsu T-R. MEMS Packaging. Institution of Engineering and Technology (IET), Herts, UK, 2004.
18  Tummala R. Fundamentals of Microsystems Packaging. McGraw Hill Professional, New York, USA, 2001.
19  Lau J, Lee C, Premachandran C, Aibin Y. Advanced MEMS Packaging. McGraw-Hill Education, New York, USA, 2009.
20  Madou MJ. Fundamentals of Microfabrication: The Science of Miniaturization, Second Edition. CRC Press, Boca Raton, FL, USA, 2002.
21  Fischer AC, Korvink JG, Roxhed N et al. Unconventional applications of wire bonding create opportunities for microsystem integration. *Journal of Micromechanics and Microengineering* 2013; **23**: 083001.
22  Milanovic V, Maharbiz M, Pister KSJ. Batch transfer integration of RF microrelays. *IEEE Microwave and Guided Wave Letters* 2000; **10**: 313–315.
23  Tejada F, Wesolek DM, Lehtonen J et al. An SOS MEMS interferometer. *Proceedings of the SPIE* 2004; **5346**: 27–36.
24  Jung IW, Krishnamoorthy U, Solgaard O. High fill-factor two-axis gimbaled tip-tilt-piston micromirror array actuated by self-aligned vertical electrostatic combdrives. *Journal of Microelectromechanical Systems* 2006; **15**: 563–571.
25  Michalicek MA, Bright VM. Flip-chip fabrication of advanced micromirror arrays. The 14th IEEE International Conference on Micro Electro Mechanical Systems; 25–25 Jan 2001; Interlaken, Switzerland; 2001: 313–316.
26  Rogge B, Moser D, Oppermann HH, Paul O, Baltes H. Solder-bonded micromachined capacitive pressure sensors. *Proceedings of the SPIE* 1998; **3514**: 307–315.
27  Waber T, Pahl W, Schmidt M et al. Flip-chip packaging of piezoresistive barometric pressure sensors. *Proceedings of the SPIE* 2013; **8763**: 87632D–87632D–8.
28  Markus KW, Dhuler VR, Roberson D et al. Smart MEMS: flip chip integration of MEMS and electronics. *Proceedings of the SPIE* 1995; **2448**: 82–92.
29  OToole E, Almeida R, Campos J et al. eWLB SiP with Sn finished passives. 2014 Electronics System-Integration Technology Conference (ESTC); 16–18 Sep 2014; Helsinki, Finland; 2014: 1–4.
30  Krohnert S, Campos J, O'Toole E. Fan-out WLP: The enabler for system-in-package on Wafer Level (WLSIP). 2012 Electronics System-Integration Technology Conference (ESTC); 17–20 Sep 2014; Amsterdam, the Netherlands; 2014: 1–8.
31  Liu F, Sundaram V, Min S et al. Chip-last embedded actives and passives in thin organic package for 110 GHz multi-band applications. 2010 Electronic Components and Technology Conference (ECTC); 1–4 Jun 2010; Las Vegas, NV, USA; 2010: 758–763.
32  Jin Y, Baraton X, Yoon SW et al. Next generation eWLB (embedded wafer level BGA) packaging. 2010 Electronics Packaging Technology Conference (EPTC); 8–10 Dec 2010; Singapore; 2010: 520–526.
33  Janus P, Grabiec P, Domanski K, Kociubinski A, Szmigiel D. Technology of hybrid integration of silicon MEMS and CMOS structures using polymer. 2008 2nd European Conference & Exhibition on Integration Issues of Miniaturized Systems - MOMS, MOEMS, ICS and Electronic Components (SSI); 9–10 Apr 2008; Barcelona, Spain; 2008: 1–4.
34  Braun T, Becker K-F, Jung E et al. Fan-out wafer level packaging for MEMS and sensor applications. Proceedings of Sensors and Measuring Systems 2014; 3–4 Jun 2014; Nuremberg, Germany; 2014: 1–5.
35  Yang HS, Bakir MS. 3D integration of CMOS and MEMS using mechanically flexible interconnects (MFI) and through silicon vias (TSV). 2010 Proceedings of Electronic Components and Technology Conference (ECTC); 1–4 Jun 2010; Las Vegas, NV, USA; 2010: 822–828.
36  Bernstein GH, Liu Q, Yan M et al. Quilt packaging: high-density, high-speed interchip communications. *IEEE Transactions on Advanced Packaging* 2007; **30**: 731–740.
37  Alper SE, Azgin K, Akin T. A high-performance silicon-on-insulator MEMS gyroscope operating at atmospheric pressure. *Sensors and Actuators A: Physical* 2007; **135**: 34–42.
38  Chen W, Chen H, Liu X, Tan X. A hybrid micro-accelerometer system with CMOS readout circuit and self-test function. *Proceedings of the SPIE* 2005; **6040**: doi:10.1117/12.664131.
39  Tilmans HAC, Raedt WD, Beyne E. MEMS for wireless communications: 'from RF-MEMS components to RF-MEMS-SiP'. *Journal of Micromechanics and Microengineering* 2003; **13**: S139.

40 Chae J, Kulah H, Najafi K. A hybrid Silicon-On-Glass (SOG) lateral micro-accelerometer with CMOS readout circuitry. The Fifteenth IEEE International Conference on Micro Electro Mechanical Systems; 24–24 Jan 2002; Las Vegas, NV, USA; 2002: 623–626.

41 Mohan A, Malshe AP, Sriram B, Natarajan K, Mohan S. Multi Chip Module (MCM) design for packaging of a MEMS pressure sensor. 6th International Conference on Polymers and Adhesives in Microelectronics and Photonics; 16–18 Jan 2007; Tokyo, Japan; 2007: 24–29.

42 Feiertag G, Winter M, Leidl A. Packaging of MEMS microphones. Proceedings of the SPIE 2009; **7362**: doi: 10.1117/12.821186.

43 Marenco N, Reinert W, Warnat S et al. Vacuum encapsulation of resonant MEMS sensors by direct chip-to-wafer stacking on ASIC. 2008 Electronics Packaging Technology Conference; 9–12 Dec 2008; Singapore; 2008: 773–777.

44 Xu G, Yan P, Chen X et al. Wafer-level chip-to-Wafer (C2W) integration of high-sensitivity MEMS and ICs. 2011 12th International Conference on Electronic Packaging Technology and High Density Packaging (ICEPT-HDP); 8–11 Aug 2011; Shanghai China; 2011: 1–5.

45 Obata S, Inoue M, Miyagi T et al. In-line wafer level hermetic packages for MEMS variable capacitor. 58th Electronic Components and Technology Conference; 27–30 May 2008; Lake Buena Vista, FL, USA; 2008: 158–163.

46 Choi WK, Premachandran CS, Xie L et al. A novel die to wafer (D2W) collective bonding method for MEMS and electronics heterogeneous 3D integration. Proceedings of 2010 Electronic Components and Technology Conference (ECTC); 1–4 Jun 2010; Las Vegas, NV, USA; 2010: 829–833.

47 Hsu Y-W, Chen J-Y, Chien H-T et al. New capacitive low-g triaxial accelerometer with low cross-axis sensitivity. Journal of Micromechanics and Microengineering 2010; **20**: 055019.

48 Prandi L, Caminada C, Coronato L et al. A low-power 3-axis digital-output MEMS gyroscope with single drive and multiplexed angular rate readout. 2011 IEEE International Solid-State Circuits Conference Digest of Technical Papers (ISSCC); 20–24 Feb 2011; San Francisco, CA, USA; 2011: 104–106.

49 Winter M, Feiertag G, Siegel C et al. Chip scale package of a MEMS microphone and ASIC stack. 23rd International Conference on Micro Electro Mechanical Systems (MEMS); 24–28 Jan 2010; Hong Kong, China; 2010: 272–275.

50 Dubuc D, De Raedt W, Carchon G et al. MEMS-IC integration for RF and millimeterwave applications. 2005 European Microwave Conference; 4–6 Oct 2005; Paris, France; doi:10.1109/EUMC.2005.1610229.

51 Oouchi A. Plastic molded package technology for MEMS sensor evolution of MEMS sensor package. 2014 International Conference on Electronics Packaging (ICEP); 23–25 Apr 2014; Toyama, Japan; 2014: 371–375.

52 Vigna B, Lasalandra E, Ungaretti T. Motion MEMS and sensors, today and tomorrow. In: Roermund AHM van, Baschirotto A, Steyaert M (eds). Nyquist AD Converters, Sensor Interfaces, and Robustness. Springer, New York, 2013: 117–127.

53 Ferraresi M, Pozzi S. MEMS sensors for non-safety automotive applications. In: Meyer G, Valldorf J, Gessner W (eds). Advanced Microsystems for Automotive Applications 2009. Springer, Berlin and Heidelberg, 2009: 355–367.

54 Chipworks. STMicroelectronics LIS331DLH 3-Axis MEMS Accelerometer – Exploratory Report, 2009; http://www.chipworks.com/TOC/STMicroelectronics_LIS331DLH_3-Axis_Accelerometer_EXR-0903-801_TOC.pdf.

55 Premachandran CS, Lau J, Xie L et al. A novel, wafer-level stacking method for low-chip yield and non-uniform, chip-size wafers for MEMS and 3D SIP applications. 58th Electronic Components and Technology Conference; 27–30 May 2008; Lake Buena Vista, FL, USA; 2008: 314–318.

56 Tian J, Sosin S, Iannacci J et al. RF–MEMS wafer-level packaging using through-wafer interconnect. Sensors and Actuators A: Physical 2008; **142**: 442–451.

57 Sugizaki Y, Nakao M, Higuchi K et al. Novel wafer-level CSP for stacked MEMS/IC dies with hermetic sealing. 58th Electronic Components and Technology Conference; 27–30 May 2008; Lake Buena Vista, FL, USA; 2008: 811–816.

58 Gibb K, Krishnamurthy R. Drilling and filling, but not in your Dentist's chair – a look at some recent history of multi-chip and through silicon via (TSV) technology. Chip Design Magazine 2008: 29–32.

59 Small M, Ruby R, Ortiz S et al. Wafer-scale packaging for FBAR-based oscillators. IEEE International Frequency Control and the European Frequency and Time Forum (FCS); 2–5 May 2011; San Francisco, CA, USA; 2011: 1–4.

60 Fujiwara T, Seki T, Sato F, Oba M. Development of RF-MEMS ohmic contact switch for mobile handsets applications. 2012 42nd European Microwave Conference (EuMC); 29 Oct–1 Nov 2012; Amsterdam, the Netherlands; 2012: 180–183.

61 VTI Technologies Oy. CMA3000 Assembly Instructions. 3 Mar 2015; http://www.muratamems.fi/sites/default/files/documents/tn68_cma3000_assembly_instructions_a.02_0.pdf.

62 Flatscher M, Dielacher M, Herndl T et al. A Bulk Acoustic Wave (BAW) based transceiver for an in-tire-pressure monitoring sensor node. IEEE Journal of Solid-State Circuits 2010; **45**: 167–177.

63 Schjølberg-Henriksen K, Taklo MMV, Lietaer N et al. Miniaturised sensor node for tire pressure monitoring (e-CUBES). In: Meyer G, Valldorf J, Gessner W (eds). Advanced Microsystems for Automotive Applications 2009. Springer, Berlin and Heidelberg, 2009: 313–331.

64 Armbruster S, Malshe AP, Lammel G et al. A novel micromachining process for the fabrication of monocrystalline Si-membranes using porous silicon. The 12th International Conference on Solid-State Sensors, Actuators and Microsystems (TRANSDUCERS 2003); 8–12 Jun 2003; Boston, MA, USA; 2003: 246–249.

65 Artmann H, Schaefer F, Lammel G et al. Monocrystalline Si membranes for pressure sensors fabricated by a novel surface micromachining process using porous silicon. Proceedings of the SPIE 2003; **4981**: doi:10.1117/12.479564.

66 Knese K, Armbruster S, Weber H et al. Novel technology for capacitive pressure sensors with monocrystalline silicon membranes. IEEE 22nd International Conference on Micro Electro Mechanical Systems (MEMS); 25–29 Jan 2009; Sorrento, Italy; 2009: 697–700.

67 Lammel G, Armbruster S, Schelling C et al. Next generation pressure sensors in surface micromachining technology. The 13th International Conference on Solid-State Sensors, Actuators, and Microsystems (TRANSDUCERS'05); 5–9 Jun 2005; Seoul, Korea; 2005: 35–36.

68 Candler RN, Park W-T, Li H et al. Single wafer encapsulation of MEMS devices. IEEE Transactions on Advanced Packaging 2003; **26**: 227–232.

69 Kim B, Melamud R, Hopcroft MA et al. Si-SiO2 composite MEMS resonators in CMOS compatible wafer-scale thin-film encapsulation. IEEE International Frequency Control Symposium, 2007 Joint with the 21st European Frequency and Time Forum; 29 May–1 Jun 2007; Geneva, Switzerland; 2007: 1214–1219.

70 Park W-T, Partridge A, Candler RN et al. Encapsulated submillimeter piezoresistive accelerometers. IEEE Journal of Microelectromechanical Systems 2006; **15**: 507–514.

71 SiTime's MEMS FirstTM Process. 3 Mar 2015. http://www.sitime.com/support2/documents/AN20001-MEMS-First-Process.pdf.

72 Lutz M, Partridge A, Gupta P et al. MEMS oscillators for high volume commercial applications. The 14th International Conference on Solid-State Sensors, Actuators and Microsystems (TRANSDUCERS'07); 10–14 Jun 2007; Lyon, France; 2007: 49–52.

73 Kiihamäki J, Dekker J, Pekko P. 'Plug-up'–a new concept for fabricating SOI MEMS devices. Microsystem Technologies 2004; **10**: 346–350.

74 Kiihamaki J, Ronkainen H, Pekko P et al. Modular integration of CMOS and SOI-MEMS using 'plug-up' concept. The 12th International Conference on Solid-State Sensors, Actuators and Microsystems (TRANSDUCERS'03); 8–12 Jun 2003; Boston, MA, USA; 2003: 1647–1650.

75 Ylimaula M, Åberg M, Kiihamaki J et al. Monolithic SOI-MEMS capacitive pressure sensor with standard bulk CMOS readout circuit. Proceedings of the 29th European Solid-State Circuits Conference; 16–18 Sep 2003; Estoril, Portugal; 2003: 611–614.

76 Smith JH, Montague S, Sniegowski JJ. Embedded micromechanical devices for the monolithic integration of MEMS with CMOS. International Electron Devices Meeting; 10–13 Dec 1995; Washington, DC, USA; 1995: 609–612.

77 Palaniapan M, Howe RT, Yasaitis J. Integrated surface-micromachined z-axis frame microgyroscope. 2002 International Electron Devices Meeting (IEDM'02); 8–11 Dec 2002; San Francisco, CA, USA; 2002: 203–206.

78 Yasaitis JA, Judy M, Brosnihan T et al. A modular process for integrating thick polysilicon MEMS devices with sub-micron CMOS. Proceedings of the SPIE 2003; **4979**: 145–154.

79 Chen TD, Kelly TW, Collins D et al. The next generation integrated MEMS and CMOS process on SOI wafers for overdamped accelerometers. The 13th International Conference on Solid-State Sensors, Actuators, and Microsystems (TRANSDUCERS'05); 5–9 Jun 2005; Seoul, Korea; **2**: 1122–1125.

80 Brosnihan TJ, Bustillo JM, Pisano AP et al. Embedded interconnect and electrical isolation for high-aspect-ratio, SOI inertial instruments. 1997 International Conference on Solid State Sensors and Actuators (TRANSDUCERS'97); 16–19 Jun 1997; Chicago, IL, USA; 1997, **1**: 637–640.

81 Lu C, Lemkin M, Boser BE. A monolithic surface micromachined accelerometer with digital output. IEEE Journal of Solid-State Circuits 1995; **30**: 1367–1373.

82 Lemkin M, Juneau T, Clark W et al. A low-noise digital accelerometer integrated using SOI-MEMS technology. International Conference on Solid-State Sensors, Actuators and Microsystems (TRANSDUCERS'99); 7–10 Jun 1999; Sendai, Japan; 1999: 1292–1297.

83 Villarroya M, Figueras E, Montserrat J et al. A platform for monolithic CMOS-MEMS integration on SOI wafers. Journal of Micromechanics and Microengineering 2006; **16**: 2203.

84 Brosnihan TJ, Brown SA, Brogan A et al. Optical IMEMS/sup /spl reg//-a fabrication process for MEMS optical switches with integrated on-chip electronics. The 12th International Conference on Solid-State Sensors, Actuators and Microsystems (TRANSDUCERS'03); 8–12 Jun 2003; Boston, MA, USA; 2003, **2**: 1638–1642.

85 Parameswaran L, Hsu C, Schmidt MA. A merged MEMS-CMOS process using silicon wafer bonding. 1995 International Electron Devices Meeting (IEDM'95); 10–13 Dec 1995; Washington, DC, USA; 1995: 613–616.

86 Guo S, Guo J, Ko WH. A monolithically integrated surface micromachined touch mode capacitive pressure sensor. *Sensors and Actuators A: Physical* 2000; **80**: 224–232.

87 Scheiter T, Kapels H, Oppermann K-G et al. Full integration of a pressure-sensor system into a standard BiCMOS process. *Sensors and Actuators A: Physical* 1998; **67**: 211–214.

88 Chau KH-L, Lewis SR, Zhao Y et al. An integrated force-balanced capacitive accelerometer for low-G applications. The 8th International Conference on Solid-State Sensors and Actuators (TRANSDUCERS'95); 25–29 Jun 1995; Stockholm, Sweden; 1995, **1**: 593–596.

89 Geen JA, Sherman SJ, Chang JF et al. Single-chip surface-micromachined integrated gyroscope with 50/spl deg//hour root Allan variance. 2002 IEEE International Solid-State Circuits Conference (ISSCC'02), Digest of Technical Papers; 7–7 Feb 2002; San Francisco, CA, USA; 2002, **1**: 426–427.

90 Wise KD, Anderson DJ, Hetke JF et al. Wireless implantable microsystems: high-density electronic interfaces to the nervous system. *Proceedings of the IEEE* 2004; **92**: 76–97.

91 Oliver AD, Baer WC, Wise KD. A Bulk-micromachined 1024-element Uncooled Infrared Imager. The 8th International Conference on Solid-State Sensors and Actuators (TRANSDUCERS'95); 25–29 Jun 1995; Stockholm, Sweden; 1995, **2**: 636–639.

92 Yoon E, Wise KD. An integrated mass flow sensor with on-chip CMOS interface circuitry. *IEEE Transactions on Electron Devices* 1992; **39**: 1376–1386.

93 DeHennis AD, Wise KD. A fully integrated multisite pressure sensor for wireless arterial flow characterization. *IEEE Journal of Microelectromechanical Systems* 2006; **15**: 678–685.

94 Ji J, Wise KD. An implantable CMOS circuit interface for multiplexed microelectrode recording arrays. *IEEE Journal of Solid-State Circuits* 1992; **27**: 433–443.

95 Yoon E, Wise KD. A wideband monolithic RMS-DC converter using micromachined diaphragm structures. *IEEE Transactions on Electron Devices* 1994; **41**: 1666–1668.

96 Ko WH, Hynecek J, Boettcher SF et al. Development of a miniature pressure transducer for biomedical applications. *IEEE Transactions on Electron Devices* 1979; **26**: 1896–1905.

97 Borky JM, Wise KD. Integrated signal conditioning for silicon pressure sensors. *IEEE Transactions on Electron Devices* 1979; **26**: 1906–1910.

98 Fedder GK. CMOS-based sensors. 2005 IEEE Sensors; 30 Oct–3 Nov 2005; Irvine, CA, USA; 2005: 125–128.

99 Ding X, Czarnocki W, Schuster JP et al. DSP-based CMOS monolithic pressure sensor for high volume manufacturing. International Conference on Solid-State Sensors, Actuators and Microsystems (TRANSDUCERS'99); 7–10 Jun, 1999; Sendai, Japan; 1999: 362–365.

100 Müller T, Brandl M, Brand O et al. An industrial CMOS process family adapted for the fabrication of smart silicon sensors. *Sensors and Actuators A: Physical* 2000; **84**: 126–133.

101 Ashruf CMA, French PJ, Sarro PM, Bressers PMMC, Kelly JJ. Electrochemical etch stop engineering for bulk micromachining. *Mechatronics* 1998; **8**: 595–612.

102 Fedder GK, Santhanam S, Reed ML et al. Laminated high-aspect-ratio microstructures in a conventional CMOS process. IEEE, The Ninth Annual International Workshop on Micro Electro Mechanical Systems, 1996, MEMS '96, Proceedings. An Investigation of Micro Structures, Sensors, Actuators, Machines and Systems; 11–15 Feb 1996; San Diego, CA, USA; 1996: 13–18.

103 Tsai JM, Fedder GK. Mechanical noise-limited CMOS-MEMS accelerometers. IEEE International Conference on Micro Electro Mechanical Systems (MEMS) 2005; 30 Jan–3 Feb 2005; Miami Beach, FL, USA; 2005: 630–633.

104 Xie H, Fedder GK. Fabrication, characterization, and analysis of a DRIE CMOS-MEMS gyroscope. *IEEE Journal of Sensors* 2003; **3**: 622–631.

105 Lakdawala H, Fedder GK. CMOS micromachined infrared imager pixel. The 11th International Conference on Solid-State Sensors and Actuators (TRANSDUCERS'01); 10–14 Jun 2001; Munich, Germany; 2001: 1548–1551.

106 Oz A, Fedder GK. CMOS/BiCMOS self assembling and electrothermal microactuators for tunable capacitors, gapclosing structures and latch mechanisms. Solid-State Sensors, Actuators and Microsystems Workshop; 6–10 Jun 2004; Hilton Head Island, SC, USA; 2004: 251–254.

107 Tseng S-H, Hung Y-J, Juang Y-Z, Lu MS-C. A 5.8-GHz VCO with CMOS-compatible MEMS inductors. *Sensors and Actuators A: Physical* 2007; **139**: 187–193.

108 Sheu M-L, Chen J-Y, Li M-S, Tsao L-J. A novel infrared microbolometer in standard CMOS-MEMS process. 2014 IEEE International Symposium on Bioelectronics and Bioinformatics; 11–14 Apr 2014; Chung Li, China; 2014: 1–4.

109 Azizi M, Sarkar N, Mansour RR. Single-chip CMOS-MEMS dual mode scanning microwave microscope. *IEEE Transactions on Microwave Theory and Techniques* 2013; **61**: 4621–4629.

110 Xie H, Erdmann L, Zhu X, Gabriel KJ, Fedder GK. Post-CMOS processing for high-aspect-ratio integrated silicon microstructures. *Journal of Microelectromechanical Systems* 2002; **11**: 93–101.

111 Tezcan DS, Eminoglu S, Akar OS, Akin T. Uncooled microbolometer infrared focal plane array in standard CMOS. *Proceedings of the SPIE* 2001; **4288**: 112–121.

112 Tepegoz M, Kucukkomurler A, Tankut F, Eminoglu S, Akin T. A miniature low-cost LWIR camera with a 160×120 microbolometer FPA. *Proceedings of the SPIE* 2014; **9070**: 90701O–90701O–8.

113 Neumann Jr. JJ, Gabriel KJ. CMOS-MEMS membrane for audio-frequency acoustic actuation. *Sensors and Actuators A: Physical* 2002; **95**: 175–182.

114 Aden JS, Bohórquez JH, Collins DM, Douglas Crook M. The third-generation HP thermal inkjet printhead. *Hewlett Packard Journal* 1994; **45**: 41–41.

115 Krause P, Obermeier E, Wehl W. Backshooter – a new smart micromachined single-chip inkjet printhead. 8th International Conference on Solid-State Sensors and Actuators (TRANSDUCERS'95); 25–29 Jun 1995; Stockholm, Sweden; 1995, **2**: 325–328.

116 Westberg D, Andersson GI. A novel CMOS-compatible inkjet head. The 9th International Conference on Solid-State Sensors and Actuators (TRANSDUCERS'97); 16–19 Jun 1997; Chicago, IL, USA; 1997; **2** : 813–816.

117 Mottin E, Bain A, Martin J-L et al. Uncooled amorphous silicon technology enhancement for 25-μm pixel pitch achievement. *Proceedings of the SPIE* 2003; **4820**: 200–207.

118 Van Kessel PF, Hornbeck LJ, Meier RE et al. A MEMS-based projection display. *Proceedings of the IEEE* 1998; **86**: 1687–1704.

119 Schmidt J-U, Friedrichs M, Bakke T et al. Technology development for micromirror arrays with high optical fill factor and stable analogue deflection integrated on CMOS substrates. *Proceedings of the SPIE* 2008; **6993**: doi:10.1117/12.787015.

120 Haspeslagha L, De Coster J, Pedreira OV et al. Highly reliable CMOS-integrated 11MPixel SiGe-based micro-mirror arrays for high-end industrial. IEEE International Electron Devices Meeting 2008 (IEDM'08); 15–17 Dec 2008; San Francisco, CA, USA; 2008: 1–4.

121 Yoneoka S, Liger M, Yama G et al. ALD-metal uncooled bolometer. 2011 IEEE 24th International Conference on Micro Electro Mechanical Systems (MEMS); 23–27 Jan 2011; Cancun, Mexico; 2011: 676–679.

122 Niklaus F, Vieider C, Jakobsen H. MEMS-based uncooled infrared bolometer arrays: a review. *Proceedings of the SPIE* 2007; **6836**: 68360D–68360D–15.

123 Sedky S, Fiorini P, Baert K, Hermans L, Mertens R. Characterization and optimization of infrared poly SiGe bolometers. *IEEE Transactions on Electron Devices* 1999; **46**: 675–682.

124 Witvrouw A, Mehta A. The use of functionally graded poly-SiGe layers for MEMS applications. *Materials Science Forum* 2005; **492–493**: 255–260.

125 Murphy DF, Ray M, Wyles J et al. Performance improvements for VOx microbolometer FPAs. *Proceedings of the SPIE* 2007; **6836**: doi:10.1117/12.755128.

126 Foulds IG, Parameswaran M. A planar self-sacrificial multilayer SU-8-based MEMS process utilizing a UV-blocking layer for the creation of freely moving parts. *Journal of Micromechanics and Microengineering* 2006; **16**: 2109.

127 Zahn JD, Gabriel KJ, Fedder GK. A direct plasma etch approach to high aspect ratio polymer micromachining with applications in bioMEMS and CMOS-MEMS. The Fifteenth IEEE International Conference on Micro Electro Mechanical Systems; 24–24 Jan 2002; Las Vegas, NV, USA; 2002: 137–140.

128 Carpentier JF, Cathelin A, Tilhac C et al. A SiGe:C BiCMOS WCDMA zero-IF RF front-end using an above-IC BAW filter. 2005 IEEE International Solid-State Circuits Conference (ISSCC), Digest of Technical Papers; 10 Feb 2005; San Francisco, CA, USA; **1**: 394–395.

129 Chipworks. InvenSense IDG-300 dual-axis angular rate gyroscope sensor, Revision 2.0. Chipworks, Ontario, Canada, 2007; http://www.chipworks.com/TOC/InvenSense_IDG-300_Dual_Axis_ARGS_MPR-0702-801_TOC.pdf.

130 Crnogorac F, Pease FRW, Birringer RP, Dauskardt RH. Low-temperature Al–Ge bonding for 3D integration. *Journal of Vacuum Science & Technology B* 2012; **30**: 06FK01.

131 Bryzek J, Nasiri S, Flannery A et al. Very large scale integration of MOEMS mirrors, MEMS angular amplifiers and high-voltage, high-density IC electronics for photonic switching. Technical Proceedings of the 2003 Nanotechnology Conference and Trade Show; Nano Science and Technology Institute, Washington, DC, USA; **2**: 428–431.

132 Bryzek J, Flannery A, Skurnik D. Integrating microelectromechanical systems with integrated circuits. *IEEE Instrumentation & Measurement Magazine* 2004; **7**: 51–59.

133 Despont M, Drechsler U, Yu R, Pogge HB, Vettiger P. Wafer-scale microdevice transfer/interconnect: its application in an AFM-based data-storage system. *Journal of Microelectromechanical Systems* 2004; **13**: 895–901.

134 Guerre R, Drechsler U, Bhattacharyya D et al. Wafer-level transfer technologies for PZT-based RF MEMS switches. *Journal of Microelectromechanical Systems* 2010; **19**: 548–560.

135 Guerre R, Drechsler U, Jubin D, Despont M. Selective transfer technology for microdevice distribution. *Journal of Microelectromechanical Systems* 2008; **17**: 157–165.

136 Javey A, SungWoo, Friedman RS, Yan H, Lieber CM. Layer-by-layer assembly of nanowires for three-dimensional, multifunctional electronics. *Nano Letters* 2007; **7**: 773–777.

137 Smith AD, Niklaus F, Paussa A et al. Electromechanical piezoresistive sensing in suspended graphene membranes. *Nano Letters* 2013; **13**: 3237–3242.

138 Smith AD, Vaziri S, Niklaus F et al. Pressure sensors based on suspended graphene membranes. *Solid-State Electronics* 2013; **88**: 89–94.

139 Mastrangeli M, Abbasi S, Varel C et al. Self-assembly from milli- to nanoscales: methods and applications. *Journal of Micromechanics and Microengineering* 2009; **19**: 083001.

140 Carlson A, Bowen AM, Huang Y, Nuzzo RG, Rogers JA. Transfer printing techniques for materials assembly and micro/nanodevice fabrication. *Advanced Materials* 2012; **24**: 5284–5318.

141 Lee SH, Chen K-N, Lu JJ-Q. Wafer-to-wafer alignment for three-dimensional integration: a review. *Journal of Microelectromechanical Systems* 2011; **20**: 885–898.

142 mCube. White Paper: The advantages of integrated MEMS to enable the Internet of moving things. 2014. http://www.mcubemems.com/wp-content/uploads/2014/06/mCube-Advantages-of-Integrated-MEMS-Final-0614.pdf.

143 Zimmer F, Lapisa M, Bakke T, Bring M, Stemme G, Niklaus F. One-megapixel monocrystalline-silicon micromirror array on CMOS driving electronics manufactured with very large-scale heterogeneous integration. *Journal of Microelectromechanical Systems* 2011; **20**: 564–572.

144 Liu CM, Chou BCS, Tsai RC-F et al. MEMS technology development and manufacturing in a CMOS foundry. The 16th International Conference on Solid-State Sensors, Actuators and Microsystems (TRANSDUCERS'11); 5–9 Jun 2011; Beijing, China; 2011: 807–810.

145 Romain Fraux. System Plus Consulting. mCube Monolithic 3-Axis Accelerometer – Reverse Costing Analysis. Nantes, France, 2013; http://www.systemplus.fr/wp-content/uploads/2013/11/S+C_RM140_mCube_3-Axis_Accelerometer_Sample1.pdf.

146 Forsberg F, Lapadatu A, Kittilsland G et al. CMOS-integrated SiGe quantum-well infrared microbolometer focal plane arrays manufactured with very large-scale heterogeneous 3-D integration. *IEEE Journal of Selected Topics in Quantum Electronics* 2015; **21**: 1–11.

147 Niklaus F, Kälvesten E, Stemme G. Wafer-level membrane transfer bonding of polycrystalline silicon bolometers for use in infrared focal plane arrays. *Journal of Micromechanics and Microengineering* 2001; **11**: 509–513.

148 Niklaus F, Jansson C, Decharat A et al. Uncooled infrared bolometer arrays operating in a low to medium vacuum atmosphere: performance model and tradeoffs. *Proceedings of the SPIE* 2007; **6542**: doi:10.1117/12.719163.

149 Forsberg F, Fischer AC, Roxhed N et al. Heterogeneous 3D integration of 17 um pitch Si/SiGe quantum well bolometer arrays for infrared imaging systems. *Journal of Micromechanics and Microengineering* 2013; **23**: 045017.

150 Forsberg F, Roxhed N, Fischer AC et al. Very large scale heterogeneous integration (VLSHI) and wafer-level vacuum packaging for infrared bolometer focal plane arrays. *Infrared physics & technology* 2013; **60**: 251–259.

151 Niklaus F, Pejnefors J, Dainese M et al. Characterization of transfer-bonded silicon bolometer arrays. *Proceedings of the SPIE* 2004; **5406**: doi:10.1117/12.565894.

152 Källhammer J-E, Pettersson H, Eriksson D et al. Fulfilling the pedestrian protection directive using a long-wavelength infrared camera designed to meet both performance and cost targets. *Proceedings of the SPIE* 2006; **6198**: doi:10.1117/12.663152.

153 Niklaus F, Haasl S, Stemme G. Arrays of monocrystalline silicon micromirrors fabricated using CMOS compatible transfer bonding. *IEEE Journal of Microelectromechanical Systems* 2003; **12**: 465–469.

154 Lapisa M, Zimmer F, Niklaus F, Gehner A, Stemme G. CMOS-integrable piston-type micro-mirror array for adaptive optics made of mono-crystalline silicon using 3-D integration. IEEE 22nd International Conference on Micro Electro Mechanical Systems (MEMS); 25–29 Jan 2009; Sorrento, Australia; 2009: 1007–1010.

155 Lapisa M, Zimmer F, Stemme G, Gehner A, Niklaus F. Drift-free micromirror arrays made of monocrystalline silicon for adaptive optics applications. *IEEE Journal of Microelectromechanical Systems* 2012; **21**: 959–970.

156 Lapisa M, Zimmer F, Stemme G, Gehner A, Niklaus F. Heterogeneous 3D integration of hidden hinge micromirror arrays consisting of two layers of monocrystalline silicon. *Journal of Micromechanics and Microengineering* 2013; **23**: 075003.

157 Ayala CL, Grogg D, Bazigos A et al. Nanoelectromechanical digital logic circuits using curved cantilever switches with amorphous-carbon-coated contacts. *Solid-State Electronics*; available in *Journal of Solid-State Electronics*, 2015.

158 Schelcher G, Fabbri F, Lefeuvre E et al. Modeling and characterization of MicroPirani vacuum gauges manufactured by a low-temperature film transfer process. *IEEE Journal of Microelectromechanical Systems* 2011; **20**: 1184–1191.

159 Matsumura T, Esashi M, Harada H, Tanaka S. Multi-band radio-frequency filters fabricated using polyimide-based membrane transfer bonding technology. *Journal of Micromechanics and Microengineering* 2010; **20**: 095027.

160 Chicherin D, Sterner M, Lioubtchenko D, Oberhammer J, Räisänen AV. Analog-type millimeter-wave phase shifters based on MEMS tunable high-impedance surface and dielectric rod waveguide. *International Journal of Microwave and Wireless Technologies* 2011; **3**: 533–538.

161 Tanaka S. Piezoelectric acoustic wave devices based on heterogeneous integration technology. 2014 IEEE International Frequency Control Symposium (FCS); 19–22 May 2014; Taipei, China; 2014: 1–4.

162 Esashi M, Tanaka S. Integrated microsystems. Advances in Science and Technology. Zurich: Trans Tech Publications Ltd. 2013; **81**: 55–64.

163 Kazior TE. Beyond CMOS: heterogeneous integration of III–V devices, RF MEMS and other dissimilar materials/devices with Si CMOS to create intelligent microsystems. *Philosophical Transactions of the Royal Society A: Mathematical, Physical & Engineering Sciences* 2014; **372**: 20130105.

164 Niklaus F, Stemme G, Lu J-Q, Gutmann RJ. Adhesive wafer bonding. *Journal of Applied Physics* 2006; **99**: 031101.

165 Niklaus F, Enoksson P, Griss P, Kalvesten E, Stemme G. Low-temperature wafer-level transfer bonding. *IEEE Journal of Microelectromechanical Systems* 2001; **10**: 525–531.

166 Niklaus F, Decharat A, Forsberg F et al. Wafer bonding with nano-imprint resists as sacrificial adhesive for fabrication of silicon-on-integrated-circuit (SOIC) wafers in 3D integration of MEMS and ICs. *Sensors and Actuators A* 2009; **154**: 180–186.

167 Stampfer C, Jungen A, Hierold C. Fabrication of discrete nanoscaled force sensors based on single-walled carbon nanotubes. *IEEE Sensors Journal* 2006; **6**: 613–617.